\documentclass[11pt,a4paper]{article}
\usepackage[dutch,english]{babel} 
\usepackage{fullpage}
\usepackage[round]{natbib} 
\usepackage{graphicx}
\usepackage{amsmath} 
\usepackage{amsfonts}
\usepackage{amssymb, amsthm}
\usepackage{bbm}

\usepackage{mathtools}

\usepackage{lscape}

\usepackage{array}
\usepackage{siunitx}
\usepackage{floatrow}
\usepackage{subcaption}
\usepackage{soul}
\usepackage{mathtools}

\usepackage{lmodern}
\usepackage{rotating}
\usepackage{algorithm2e}

\usepackage{tikz}
\usepackage{tabularx}
\usepackage{multirow}
\usetikzlibrary{automata,positioning}
\usepackage{dcolumn}
\usepackage{subcaption}
\usepackage{enumitem}
\usepackage{abstract}
\usepackage[left=2.8cm,right=2.8cm,top=2.8cm,bottom=2.8cm,twoside]{geometry}
\usepackage{geometry}
\usepackage{setspace}
\usepackage[hidelinks]{hyperref} 
\usepackage{ragged2e}
\usepackage{float}
\usepackage{colortbl}
\usepackage[flushleft]{threeparttable}
\usepackage{booktabs}
\usepackage{caption}
\usepackage{booktabs,threeparttable}
\usepackage{placeins}
\usepackage{caption}
\usepackage{breqn}
\usepackage{verbatim}
\usepackage{eurosym}
\usepackage{changepage}
\usepackage{listings}
\usepackage{xcolor}
\usepackage{fancyvrb}

\usepackage{color} 
\definecolor{mygreen}{RGB}{28,172,0} 
\definecolor{mylilas}{RGB}{170,55,241}

\DeclareSIUnit{\veh}{veh}
\DeclareMathOperator*{\argmax}{\arg\!\max}

\newtheorem{standing assumption}{Standing Assumption}

\newcommand{\mc}{\mathcal}

\newcommand{\R}{\mathds{R}}

\usepackage{amsmath} 
\usepackage{amssymb}  
\usepackage{hyperref}
\usepackage{dsfont}
\usepackage{xcolor}
\usepackage{mathtools}
\usepackage{enumerate}
\usepackage[symbol*]{footmisc}

\usepackage{tikz,pgfplots}

\usetikzlibrary{arrows.meta}

\usepackage{multirow}
\usepackage{makecell}
\usepackage{adjustbox}
\usepackage{optidef}

\usepackage{thmtools,thm-restate}

\begin{document}
\LARGE \noindent Gaussian processes for real-time urban traffic estimation using a fleet of drones

\normalsize

\begin{flushleft}
\noindent Batuhan Avci$^{*1}$, Marko Susnjar\footnote[1]{These authors contributed equally to this work.}$^{1}$, Marko Maljkovic$^{1}$, Nikolas Geroliminis$^{1}$ \\
\scriptsize
$^1$ Urban Transport Systems Laboratory (LUTS),
\'{E}cole Polytechnique F\'{e}d\'{e}rale de Lausanne (EPFL), Switzerland \\

Contact: batuhan.avci@epfl.ch,
marko.susnjar@epfl.ch,
marko.maljkovic@epfl.ch,
nikolas.geroliminis@epfl.ch
\end{flushleft}

\noindent 
\color{black}
\\
\noindent\rule[0.5ex]{\linewidth}{1pt}

\normalsize \noindent \textbf{Abstract}\\ \small
Real-time urban traffic monitoring is essential for modern city management, yet conventional static sensors provide only limited spatial coverage. Autonomous drone fleets can complement this infrastructure by offering flexible, high-resolution mobile sensing across different parts of the network. This paper introduces a spatio-temporal Gaussian process (GP) framework for reconstructing network-wide traffic states from partial aerial observations. The model accounts for the physical structure of the road network through a graph Matérn kernel, which embeds network topology into the spatial covariance structure, and captures temporal dependencies through an exponential temporal kernel. Within this framework, historical loop detector data provide an informative prior mean that is updated in real time using incoming drone observations, yielding network-wide point estimates together with predictive uncertainty. We validate our framework in two distinct case studies: a microscopic simulation of Barcelona and real-world drone data from Athens. The results demonstrate that our approach outperforms historical baselines in estimating and predicting Vehicle Hours Traveled (VHT), notably reducing the required fleet size. Furthermore, leveraging the model’s predictive uncertainty to guide drone movements yields additional performance gains, highlighting the potential for future uncertainty-aware fleet routing algorithms.

\\
\noindent\rule[0.5ex]{\linewidth}{1pt}

\clearpage
\section{Introduction}

The ability to observe traffic conditions across an urban road network in real time is a fundamental requirement for modern traffic management systems. These conditions are typically described by traffic variables such as speed, density, and flow, and estimating them over a road network becomes crucial for various traffic applications like adaptive signal control (\cite{tsitsokas2023}; \cite{li2024cooperative}; \cite{zoabi2026}) or perimeter control (\cite{hamedmoghadam2022percolation}; \cite{kouvelas2023linear}; \cite{moshahedi2023}). However, directly observing traffic variables across an entire urban road network is often infeasible, as it is not possible to monitor an entire city simultaneously. Consequently, the network-wide state must be inferred from partial and noisy observations gathered at the current and previous time steps. The difficulty is compounded by the structure of the problem. A network contains thousands of road segments whose states are tightly coupled through intersections, and conditions on each segment change continuously under signal cycles, demand fluctuations, and incidents. Any estimation method must respect this coupling rather than treat segments in isolation.

\medskip Conventional traffic monitoring typically relies on loop detectors and cameras installed at fixed positions along the roads. Loop detectors are widely deployed and relatively accessible, but their simple measurement mechanism makes them particularly susceptible to noise and local disturbances. Cameras, on the other hand, can provide richer visual information, yet they often require complex software calibration and are sensitive to environmental noise like occlusion and weather. Despite these differences, both sensing technologies share a fundamental limitation - their static placement within the road network. Once installed, they cannot be repositioned in response to evolving traffic patterns, incidents, or temporary demand shifts. Over time, their spatial coverage may become misaligned with the parts of the network where information is most needed, leaving some areas unobserved while others are monitored redundantly. Thus, conventional fixed sensors typically provide only localized and limited information, even at the link level (\cite{xiong2025}). 
 
\medskip Drones have recently emerged as a flexible alternative for traffic monitoring~(\cite{elloumi2018, barmpounakis2020, maljkovic2025}). Acting as a fleet of mobile sensors, drones can be directed to an area of interest and, through modern computer vision techniques, produce high-resolution traffic variables for every road segment within their field of view~(\cite{khan2017uavtrajectories, zhou2026methodological}). The pNEUMA experiment demonstrated this capability at scale, using a fleet of low-cost, camera-equipped, commercial drones to record traffic over a congested district of Athens and to extract individual vehicle trajectories across the entire monitored area~(\cite{barmpounakis2020}). Subsequent advances in trajectory extraction from aerial footage have improved the accuracy and geo-referencing of these reconstructions~(\cite{FONOD2025105205}), allowing us to treat each drone as a mobile sensing unit capable of estimating the traffic variable of interest through onboard computation. As a result, these developments have supported a wide range of downstream use cases, including emission and noise estimation~(\cite{espadaler2023}), real-time traffic flow estimation~(\cite{ke2019}), outdoor parking monitoring~(\cite{Kim2024}), and driving-behavior analysis~(\cite{krajewski2018}). However, the main limitation remains the spatial extent that can be covered at any given time. Even a fleet of drones can cover only a fraction of a large network at any given instant, and as drones move along their trajectories, the observed subset of segments changes from one time step to the next. Therefore, network coverage is time-varying by construction.

\begin{figure}[t]
    \centering
    \includegraphics[width=\textwidth]{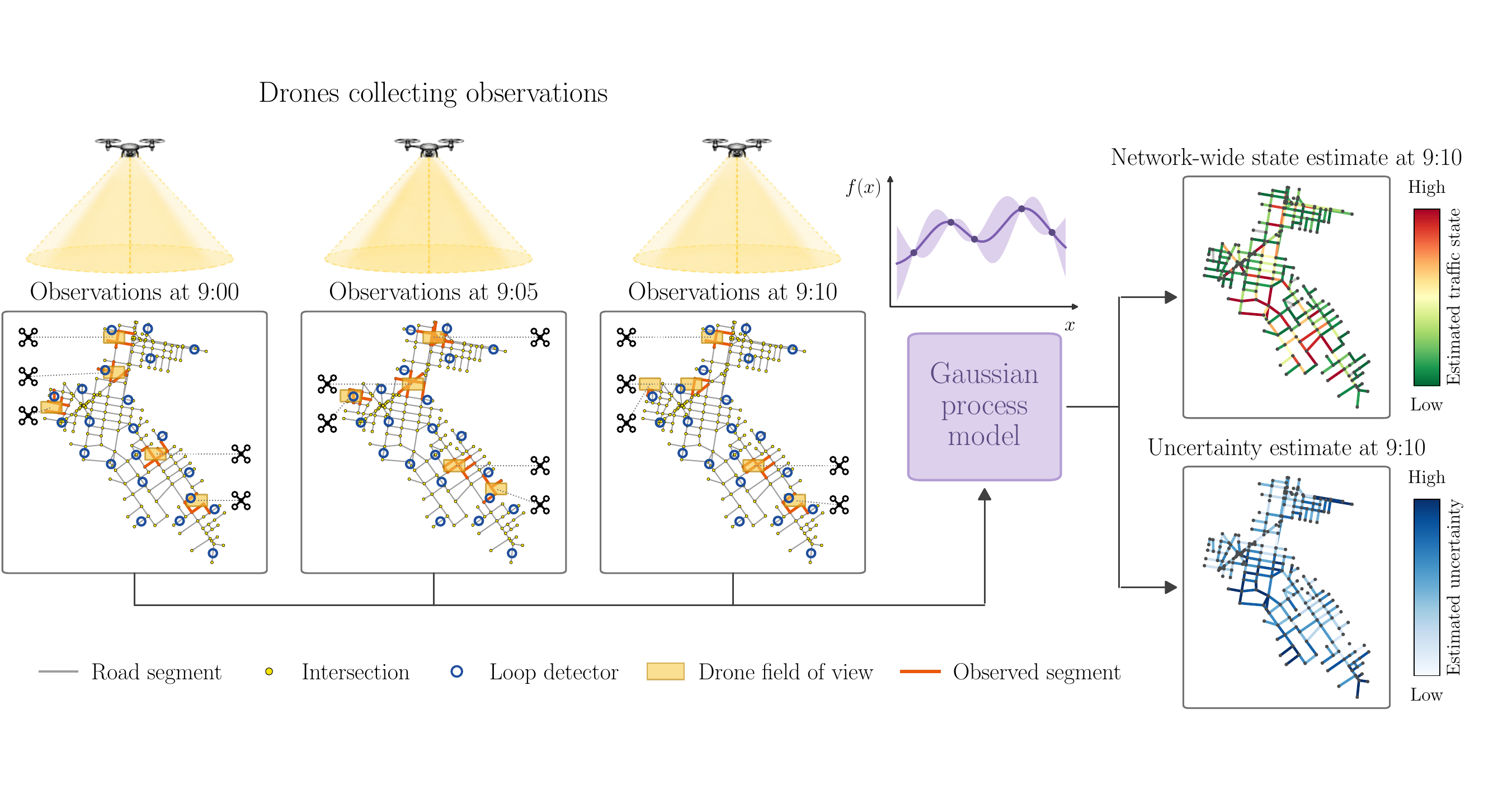}
    \caption{Overview of the proposed framework: fleet of drones collects partial traffic observations over consecutive time steps, and the Gaussian process model produces network-wide traffic state and uncertainty estimates.}
    \label{fig:teaser}
\end{figure}

\medskip This partial observability defines the central challenge we aim to address in this work: how can traffic variables be recovered at the network scale when only noisy measurements from a small and dynamically changing subset of segments are available? This setting naturally motivates an estimator that can propagate information from observed to unobserved segments while also representing the reliability of its estimates. Since the observations are collected over a road network, this inference should also account for its physical structure so that the spatial correlations in the estimated traffic state reflect the underlying topology rather than merely geometric proximity. These considerations point toward a probabilistic and structure-aware modeling framework. Gaussian Process (GP) regression is well suited for this task, as it provides both a predictive mean and a measure of predictive uncertainty at unobserved locations (\cite{Rasmussen2006Gaussian}). These uncertainty estimates could be particularly valuable in the present setting because they could be used not only to assess the reliability of the reconstructed traffic state, but also to guide downstream decisions about where the drone fleet should collect information next. 


\medskip In this work, we propose a spatio-temporal GP framework for inferring traffic variables from partial observations collected by a drone fleet, as illustrated in Figure~\ref{fig:teaser}. The road network is represented as a weighted graph, with traffic variables jointly modeled as a sample from a GP. At each discrete time step, observations are available only at a subset of nodes determined by the drones' instantaneous positions. Based on these partial measurements, the GP produces posterior estimates of the traffic variables over the entire network, together with associated uncertainty quantification even for unobserved segments. We train and evaluate the proposed model on two case studies, a microscopic Aimsun simulation of the city of Barcelona (\cite{xiong2025}) and real-world drone data collected in the city of Athens in the pNEUMA experiment (\cite{barmpounakis2020}).

\medskip The remainder of the paper is organized as follows. Section~\ref{sec:relwork} reviews related work and positions our contribution. Section~\ref{sec:method} presents the methodology, including the road-network model and the spatio-temporal covariance structure. Section~\ref{sec:data} describes the two datasets and the training procedure. Section~\ref{sec:results} reports and discusses the numerical results for both case studies, together with the uncertainty analysis of proposed framework. Section~\ref{sec:con} concludes and outlines directions for future work.

\section{Related work}\label{sec:relwork}

The problem we consider sits at the intersection of three lines of research: drone-based mobile sensing, data-driven traffic forecasting, and probabilistic regression with Gaussian processes. These areas share the common goal of recovering traffic conditions from incomplete observations, but differ in their assumptions about the sensing infrastructure, the availability of historical data, and the type of predictions they provide. 

\medskip Existing literature has demonstrated the usefulness of unmanned aerial vehicles for vehicle detection, tracking, counting, trajectory extraction, and the estimation of traffic variables such as speed, density, flow, and queue length~(\cite{kanistras2013uavsurvey, khan2017uavframework, wang2016uavtracking, khan2018uavanalysis}). Other studies consider the deployment or routing of drones for traffic monitoring under coverage, mobility, and operational constraints~(\cite{chow2016dynamicuav}). Taken together, these studies establish drones as valuable mobile sensors for traffic applications. However, they mainly focus either on extracting traffic information from the area currently observed by the drone or on planning future sensing locations. In contrast, our focus is on the network-level inference problem that arises once drone measurements are available, i.e., how to use sparse aerial observations to reconstruct and predict traffic states at both observed and unobserved road segments. 

\medskip Most modern data-driven traffic forecasting methods rely on deep neural architectures that learn spatial and temporal patterns from large volumes of historical data. Graph neural networks (GNNs) are particularly natural for this task because road networks are inherently graph-structured. Foundational graph-learning models~(\cite{defferrard2016, kipf2017, hamilton2017, velickovic2018, gilmer2017}) have inspired traffic-specific architectures that apply a graph operator over a fixed road graph, together with a component that captures temporal dynamics, such as the Diffusion Convolutional Recurrent Neural Network (DCRNN)~(\cite{li2018dcrnn}) and the Spatio-Temporal Graph Convolutional Network (STGCN)~(\cite{yu2018stgcn}). Later models relax the assumption that the physical road topology fully captures traffic dependence. Graph WaveNet learns the graph structure from data directly~(\cite{wu2019gwn}) rather than fixing it in advance, while D2STGNN allows dynamic spatial-temporal dependencies to evolve over time~(\cite{shao2022d2stgnn}). Transformer-based methods~(\cite{vaswani2017}) provide an alternative powerful family of models that uses attention mechanisms to capture dependencies across road segments and time steps ~(\cite{ xu2020sttransformer, ye2022mgt, shao2022step, jiang2023pdformer, lee2024testam}). These models effectively capture spatio-temporal correlations when sufficiently rich training data are available. However, many commonly used GNN- and transformer-based architectures are primarily designed for point estimates rather than predictive distributions, and treat the road graph as a deterministic structural input rather than a probabilistic prior that could be exploited when observations are sparse.

\medskip On the other hand, Gaussian processes (GPs) provide an alternative probabilistic framework for traffic estimation and prediction. While deep neural networks require vast datasets to generalize effectively, GP models can achieve strong predictive performance with significantly smaller volumes of training data. Furthermore, unlike standard point-forecasting models, they provide both a predictive mean and a posterior uncertainty~(\cite{Rasmussen2006Gaussian}) at any given location and time. This uncertainty can be especially useful in mobile sensing, since it identifies regions where the estimate is least reliable and can therefore guide future sensing decisions. The connection between GPs and informative sensing has been studied in sensor placement and active learning, including near-optimal sensor placement~(\cite{krause2008}), no-regret Bayesian optimization with GP models~(\cite{srinivas2010}), and time-varying extensions for nonstationary settings~(\cite{bogunovic2016}). Moreover, they have also been used for traffic state estimation. For example, \cite{chen2012} model traffic as a spatio-temporal GP and fuse observations from mobile sensors, while \cite{chen2015} extend related ideas to mobility-on-demand systems. More recently, \cite{wu2024aniso} propose anisotropic GPs for traffic state estimation from sparse trajectory data. Although these works share our goal of inferring traffic conditions from partial observations, they typically define covariance in Euclidean space or rely on trajectory-based measurements. In road networks, Euclidean distance can be an erroneous proxy for traffic dependence as two segments may be geographically close but weakly connected, while more distant segments may be strongly coupled through the network topology. This motivates investigating covariance functions defined directly on graphs. In particular, graph Mat\'{e}rn kernels provide a way to construct Gaussian processes whose smoothness and correlation structure are induced by the graph Laplacian~(\cite{borovitskiy21a}).

\medskip The preceding literature highlights the need for models that combine mobile drone observations, road-network structure, and uncertainty-aware prediction. One natural way to achieve this integration is by combining fixed and mobile sensing sources. For example, \cite{wright2016fusing} combine loop detector and GPS probe measurements in a particle filter to estimate freeway density, while \cite{zockaie2018nfd} allocate a limited sensing budget between detector locations and sampled probe trajectories to estimate the network fundamental diagram. Our work shares this multi-source structure, with drones taking the role of the mobile sensor. In this sense, the most closely related study is \cite{xiong2025}, which forecasts urban traffic speed by combining drone observations with loop detector data. Although both approaches are multi-source, drone-enabled, and graph-based, they address different estimation problems and differ in two key respects. First, while \cite{xiong2025} employs a neural architecture to learn a deterministic mapping from chosen traffic features to link-level speeds, we formulate network-wide traffic estimation as probabilistic inference, obtaining both posterior estimates and their associated uncertainty. Second, the two approaches incorporate the graph structure differently. In \cite{xiong2025}, the topology guides learned message passing between neighboring road segments, whereas in our framework it directly defines a graph Mat\'{e}rn covariance prior that provides an explicit probabilistic model of how traffic information is correlated and propagated across the network.

\medskip With that in mind, our main contributions are threefold and can be summarized as follows. First, we formulate network-wide traffic state estimation and one-step-ahead prediction from partial observations as inference in a spatio-temporal Gaussian process, using a graph Mat\'{e}rn spatial kernel to encode road topology and an exponential temporal kernel to capture temporal dependence. Loop detector data are used to construct an informative prior mean, while drone observations update the posterior distribution in real time. Second, we validate the framework in two complementary settings: a microscopic Aimsun simulation of Barcelona and real drone data from Athens. In both cases, we compare static and dynamic GP-based fleet deployments against historical-mean baselines. Third, beyond prediction accuracy, we assess predictive uncertainty at unobserved segments using prediction-interval metrics and discuss how this uncertainty can inform future drone-sensing decisions.

\section{Methodology}\label{sec:method}

GPs extend the concept of a multivariate Gaussian distribution to stochastic processes, providing a probabilistic framework for learning unknown functions while quantifying predictive uncertainty (\cite{Rasmussen2006Gaussian}). A GP can be viewed as a distribution over functions $f:\mc X\rightarrow\R$, fully characterized by its mean function $m:\mc X\rightarrow\R$ and covariance (i.e., kernel) function $k:\mc X^2\rightarrow\R$. In this framework, the latent function $f$ is typically modeled as a sample from a GP denoted as
\begin{equation} \label{eq:gp_preliminaries}
    f(\mathbf{x}) \sim \mathcal{GP}(m(\mathbf{x}), k(\mathbf{x}, \mathbf{x}')),
\end{equation}
where $\mathbf{x}, \mathbf{x}' \in \mathcal{X}$ represent $d$-dimensional feature vectors encapsulating specific attributes within the input domain. For any finite collection of inputs $\mathbf{X} = \{\mathbf{x}_1, \dots, \mathbf{x}_N\}$, the corresponding function evaluations $\mathbf{f} = [f(\mathbf{x}_1), \dots, f(\mathbf{x}_N)]^\top$ follow a joint Gaussian distribution $\mathbf{f} \sim \mathcal{N}(\mathbf{m}, \mathbf{K})$, where $\mathbf{m}=[m(\mathbf{x}_1), \dots, m(\mathbf{x}_N)]^T$ represents the prior mean vector and the covariance matrix elements are given by $\mathbf{K}_{i,j} = k(\mathbf{x}_i, \mathbf{x}_j)$. This formulation enables the model to provide not only point estimates but also a principled measure of predictive uncertainty through the variance of the distribution. 

In our setting, traffic variables are indexed by road segments and time, so $\mc X$ is a product of a discrete set of road segments with a temporal horizon. Since segments interact primarily through the connections they share, we take the road network itself as the spatial domain and define the covariance directly over its topology, so that correlations follow the connectivity of the network. We formalize this construction below, first defining the road network model and then the spatio-temporal covariance
structure built on top of it.


\subsection{Road network model} \label{ss:road_network}
 
We represent the urban road network as an undirected weighted graph $\mathcal{G} = (\mathcal{V}, \mathcal{E}, \mathbf{W})$, where $\mathcal{V} = \{v_1, \dots, v_n\}$ is a set of $n$ nodes and $\mathcal{E} \subseteq \mathcal{V} \times \mathcal{V}$ is the set of edges. Each node $v_i \in \mathcal{V}$ represents the midpoint of a road segment. An edge $(v_i, v_j) \in \mathcal{E}$ exists if and only if the corresponding road segments are physically connected in the network. This graph representation reflects the fact that traffic states evolve over a structured topology, where interactions arise primarily between adjacent segments linked through intersections or direct continuation of the roadway. 

\medskip Although urban road networks are inherently directional, we model $\mathcal{G}$ as undirected since the underlying graph is used to encode spatial dependence, not to explicitly describe vehicle movement. In this context, an edge indicates that two road segments are topologically adjacent and may exhibit correlated traffic states. These dependencies need not follow only the nominal direction of travel as congestion on a downstream link can affect upstream conditions through spillback, while information about congestion may also influence drivers’ route choices and therefore traffic conditions on neighboring links. Therefore, representing adjacent segments as mutually coupled provides a practical approximation for capturing local spatial correlations and allows us to later adopt well-understood undirected graph kernels. 

The strength of these spatial interactions is encoded by the symmetric weight matrix $\mathbf{W} \in \mathbb{R}_{\geq 0}^{n \times n}$, whose entries $W_{ij}$ quantify the pairwise coupling strength between nodes $v_i$ and $v_j$. Specifically, we define
\begin{equation}
    W_{ij} = 
    \begin{cases} 
        \omega(v_i, v_j) > 0 & \text{if } (v_i, v_j) \in \mathcal{E}, \\
        0 & \text{otherwise,}
    \end{cases}
\end{equation}
with $W_{ii} = 0$ for all $i$ (i.e., no self-loops) and where $\omega: \mathcal{V} \times \mathcal{V} \to \mathbb{R}_{>0}$ is a weighting function. In the context of urban traffic, this weighting is intended to capture the intuition that directly connected segments that are geographically close tend to exhibit stronger dependence, since congestion, and flow propagation are more likely to affect neighboring parts of the network than distant ones. A natural choice for $\omega$ could be the inverse of the distance between segment midpoints, and is used in this paper.
 
\medskip This graph construction serves as the spatial backbone of the GP-based traffic model introduced in the following subsection. By defining $\mathcal{G}$ over road-segment midpoints, the model captures how local traffic states interact through the underlying road topology. Each realization of traffic on $\mathcal{G}$ thus corresponds to a network-wide traffic state, whose spatial correlations are shaped by the connectivity and geometry of the underlying traffic infrastructure.

\subsection{Problem definition}
 
Let $\mathcal{X} = \mathcal{V} \times \mathcal{T}$ denote the spatio-temporal input domain, where $\mathcal{V}$ is the node set of the road network graph $\mathcal{G}$ and $\mathcal{T} \subseteq \mathbb{R}_{\geq 0}$ is the temporal horizon. Each input $\mathbf{x} = (v, t) \in \mathcal{X}$ indexes a spatial location $v \in \mathcal{V}$ at time $t \in \mathcal{T}$.
\medskip

For a particular traffic variable (flow, density, speed, etc.), we model it on a network-wide level as a function $f: \mathcal{X} \to \mathbb{R}$, drawn from a spatio-temporal GP
\begin{equation}\label{eq:trafficGP} 
    f \sim \mathcal{GP}\left(m,\; k\right), \qquad m: \mathcal{X} \to \mathbb{R}, \quad k: \mathcal{X} \times \mathcal{X} \to \mathbb{R},
\end{equation}
where $m(\mathbf{x}) = m((v,t))$ is the expected mean value of the traffic variable on road segment $v\in\mc V$ at time step $t\in\mc T$, and $k(\mathbf{x}, \mathbf{x}')$ is the kernel encoding spatio-temporal dependencies induced by the network topology and temporal dynamics.
 
\medskip We consider a mobile sensing framework comprising a fleet of $P$ drones that collectively monitor the road network over a set of $L$ discrete equidistant observation time steps $\{t_j\}_{j=1}^{L} \subset \mathcal{T}$. At each time step $t_j$, a drone $p \in \{1, \dots, P\}$ observes a subset of road links $\mathcal{V}_j^{(p)} \subseteq \mathcal{V}$ determined by its instantaneous position. The aggregate set of observed nodes at time $t_j$ is then
\begin{equation}
    \mathcal{V}_j = \bigcup_{p=1}^{P} \mathcal{V}_j^{(p)}\:\:\:\text{with}\:\:\: n_j = |\mathcal{V}_j|.
\end{equation}
Both $\mathcal{V}_j^{(p)}$, and therefore $\mathcal{V}_j$, may vary across time steps depending on the sensing policy. For instance, a hovering drone yields a static $\mathcal{V}_j^{(p)}$, whereas a cruising trajectory produces a time-varying coverage pattern. The observation of a road segment $v_i \in \mathcal{V}_j$ at time $t_j$ is modeled as
\begin{equation}
    y_{ij} = f(v_i, t_j) + \varepsilon_{ij}, \qquad \varepsilon_{ij} \overset{\text{i.i.d.}}{\sim} \mathcal{N}(0, \sigma_n^2)
\end{equation}
where $\sigma_n^2$ denotes the noise variance, capturing both sensor measurement noise and the uncertainty arising from the reconstruction and identification of individual vehicle trajectories. Across the monitoring horizon comprising all $L$ time steps, the total number of collected observations is $N = \sum_{j=1}^{L} n_j$. We enumerate these as $\mathbf{x}_1, \dots, \mathbf{x}_N \in \mathcal{X}$, where each $\mathbf{x}_s = (v_i, t_j)$ corresponds to a specific node $v_i \in \mathcal{V}_j$ observed at time $t_j$, and denote
\begin{equation}
    \mathbf{X} = \{\mathbf{x}_1, \dots, \mathbf{x}_N\} \subset \mathcal{X}, \qquad \mathbf{y} = [y_1, \dots, y_N]^\top \in \mathbb{R}^N
\end{equation}
where $y_s$ is the noisy observation associated with input $\mathbf{x}_s$.
 
\medskip Since $f$ is modeled as a GP and the observation noise is assumed to be Gaussian, the posterior $f \mid \mathbf{X}, \mathbf{y}$ is also a GP. Hence, the posterior mean and covariance for arbitrary query points $\mathbf{x}_* = (v_*, t_*) \in \mathcal{X}$ and $\mathbf{x}_*' = (v_*', t_*') \in \mathcal{X}$ are given by
\begin{align}
    m(\mathbf{x}_* \mid \mathbf{X}, \mathbf{y}) &= m(\mathbf{x}_*)+\mathbf{k}_{*}^\top \left(\mathbf{K}_{\mathbf{XX}} + \sigma_n^2 \mathbf{I}_N\right)^{-1} (\mathbf{y} - \mathbf{m})\label{eq:post_mean} \\
    \mathrm{cov}(\mathbf{x}_*, \mathbf{x}_*' \mid \mathbf{X}, \mathbf{y}) &= k(\mathbf{x}_*, \mathbf{x}_*') - \mathbf{k}_{*}^\top \left(\mathbf{K}_{\mathbf{XX}} + \sigma_n^2 \mathbf{I}_N\right)^{-1} \mathbf{k}_{*'} \label{eq:post_cov}
\end{align}
where $\mathbf{m}=[m(\mathbf{x}_s)]_{s=1}^N\in\R^N$ is the prior mean vector, $[\mathbf{K}_{\mathbf{XX}}]_{ss'} = k(\mathbf{x}_s, \mathbf{x}_{s'})$ is the $N \times N$ covariance matrix over the observed inputs, $\mathbf{k}_{*} = [k(\mathbf{x}_*, \mathbf{x}_s)]_{s=1}^{N} \in \mathbb{R}^N$ is the cross-covariance vector between the query point and the observations, and $\mathbf{I}_N$ is the $N$-dimensional identity matrix. The predictive variance at $\mathbf{x}_*$ is obtained by setting $\mathbf{x}_*' = \mathbf{x}_*$ in~\eqref{eq:post_cov}.

\subsection{Spatio-temporal covariance structure}\label{sec:kernel}
 
The kernel $k$ must jointly encode spatial correlations governed by the road network topology and temporal correlations capturing the dynamics of intra-day traffic fluctuations. We use a separable (product) kernel, factoring the joint covariance into independent spatial and temporal components
\begin{equation}\label{eq:sep_kernel}
    k(\mathbf{x}, \mathbf{x}') = \sigma_f^2 \; k_{\mathcal{G}}(v, v') \; k_{\mathcal{T}}(t, t')
\end{equation}
where $\mathbf{x} = (v,t)$, $\mathbf{x}' = (v',t')$, and parameter $\sigma_f^2 > 0$ is the signal variance (i.e., output scale). Note that the separability assumption implies that the spatial correlation structure is time-invariant, and, conversely, that the temporal correlation structure does not depend on the spatial location.
 
\medskip For the spatial component $k_{\mathcal{G}}$ in~\eqref{eq:sep_kernel}, we employ the Graph Mat\'{e}rn kernel (\cite{borovitskiy21a}), a discrete analogue of continuous Mat\'{e}rn fields defined by the graph Laplacian $\boldsymbol{\Delta} = \mathbf{D} - \mathbf{W}$, where $\mathbf{D} = \operatorname{diag}(d_1, \dots, d_n)$ with $d_i = \sum_{j} W_{ij}$ being the degree matrix. The Laplacian admits the eigendecomposition $\boldsymbol{\Delta} = \mathbf{U} \boldsymbol{\Lambda} \mathbf{U}^\top$, with orthogonal eigenvectors $\mathbf{U} = [\mathbf{u}_1, \dots, \mathbf{u}_n]$ and eigenvalues $0 \leq \lambda_1 \leq \lambda_2 \leq \cdots \leq \lambda_n$. Following the characterization of Mat\'{e}rn kernels on graphs, the spatial covariance matrix is defined as
\begin{equation}\label{eq:graph_matern}
    \mathbf{K}_{\mathcal{G}} = \left( \frac{2\nu_g}{\kappa_g^2} \mathbf{I}_n + \boldsymbol{\Delta} \right)^{-\nu_g} = \mathbf{U} \operatorname{diag}\!\left(\left\{ \left( \frac{2\nu_g}{\kappa_g^2} + \lambda_l \right)^{-\nu_g} \right\}_{l=1}^{n}\right) \mathbf{U}^\top
\end{equation}
where $\nu_g > 0$ and $\kappa_g > 0$ are the smoothness and lengthscale parameters, respectively. The element-wise covariance between nodes $v$ and $v'$ is
\begin{equation}\label{eq:graph_matern_elem}
    k_{\mathcal{G}}(v, v') = \sum_{l=1}^{n} \left( \frac{2\nu_g}{\kappa_g^2} + \lambda_l \right)^{-\nu_g} u_l(v) \, u_l(v')
\end{equation}
where $u_l(v)$ denotes the $v$-th component of the $l$-th eigenvector. By construction, the kernel assigns high correlation to nodes that are topologically close on $\mathcal{G}$, ensuring that information propagates strictly through valid network connections.
 
\medskip For the temporal component $k_{\mathcal{T}}$ in~\eqref{eq:sep_kernel}, we want to capture the dynamics of traffic evolution along the time axis. In urban traffic, changes arise from congestion waves, signal cycles, and incidents. To model this behavior, we use the standard Mat\'{e}rn kernel with smoothness parameter $\nu = 1/2$, which reduces to the exponential kernel
\begin{equation}\label{eq:temporal_kernel}
    k_{\mathcal{T}}(t, t') = \exp\!\left( -\frac{|t - t'|}{\ell_{\mc T}} \right)
\end{equation}
where $\ell_{\mc T} > 0$ is the temporal lengthscale governing the rate of correlation decay. Larger values of $\ell_{\mc T}$ encode stronger persistence, while smaller values allow to track fluctuations.

\section{Data and training}\label{sec:data}

The methodology described in Section~\ref{sec:method} was applied to data from simulations in the city of Barcelona and to data collected in real world using a fleet of drones in the city of Athens. In this section, we describe both datasets and the procedure used to train the GP and estimate its parameters from the available observations.

\subsection{Simulation data overview}\label{sec:data_sim}

The first dataset used in this paper was generated by (\cite{xiong2025}) using the microscopic traffic simulator Aimsun (\cite{casas2010traffic}), which provides trajectories for each vehicle, together with complete information on the road network topology and geometry of road segments. The simulated network corresponds to the urban area in the city of Barcelona with 1570 road segments and 721 intersections, modeled as an undirected graph with 1570 nodes and 2803 edges connecting the road segments that have an intersection in common (as described in Section \ref{sec:method}).

\medskip
Traffic simulators offer the possibility to generate large datasets needed for stable training and robust testing  performance, which is often an issue in real world experiments. A total of 101 time series were generated from simulation runs with varying traffic demand levels. Each simulation was initialized using historically calibrated origin-destination data for a specific time period of the day, which was then randomly perturbed and multiplicatively scaled to produce augmented, but realistic demand profiles. The original historic demand corresponds to the scale $1.0$, and additional 100 series were generated with demand scales in $\{1.20, 1.35, 1.50, 1.65, 1.80\}$, thereby representing a range of increasingly congested traffic scenarios in the urban area. Every time series captures a four-hour interval of a single day, consisting of a 15-minute warm up period, 1 hour and 45 minutes of main demand, followed by two hours without new vehicle generation to allow the network to unload. Only the two-hour interval containing realistic traffic conditions, beginning immediately after the warm-up phase and together with the first 15 minutes of the unload phase, is retained for analysis.

\medskip
The simulator provides precise vehicle trajectories characterized by the vehicle positions at each simulation time step and the exact times at which vehicles enter and exit individual road segments. From these trajectories, segment-level traffic variables, such as Vehicle Kilometers Traveled (VKT), Vehicle Hours Traveled (VHT), speed, and related quantities, can be computed. Importantly, the same trajectory representation can also be obtained from real-world drone observations. As demonstrated in (\cite{FONOD2025105205}), in well-performed stable drone flights, computer vision techniques can accurately reconstruct the trajectories of the majority of vehicles within the monitored area. Accordingly, we assume that whenever a road segment is observed by a drone, the ground-truth value of the traffic variable of interest on that segment is available. 

\medskip Since loop detectors are often already installed at fixed locations in practice, their measurements provide a natural source of information to incorporate into the estimation framework. To simulate such fixed-point sensing infrastructure, a virtual loop detector is placed at the midpoint of each road segment, following the setup in~\cite{xiong2025}. Each virtual loop detector is modeled as a single virtual line that provides local flow and velocity measurements by counting vehicles as they cross it and estimating their speeds from the distance traveled between the consecutive time steps immediately before and after detection. However, these measurements are inherently localized and may not fully represent the average velocity over an entire road segment. This limitation becomes particularly relevant when traffic conditions vary along the segment, for instance due to queues, stop-and-go waves, or localized congestion near intersections. Consequently, quantities such as velocity and VHT estimates may be more sensitive to this measurement approximation, resulting in a discrepancy of 40-50\% on average, while the VKT estimates are expected to be less affected and can be deemed a good approximation of the ground-truth values. This setup therefore emulates the type of noise and measurement errors that would arise in real-world loop detector data.


\medskip When vehicle origins and destinations are predefined, the total VKT is predominantly determined by the distance the vehicles travel rather than the duration of their delays. Therefore, VKT alone does not directly reflect the intensity of congestion. Since congestion mainly increases travel time, VHT provides a more informative metric, as it scales when vehicles spend more time in the network due to lower speeds, queues, or stop-and-go traffic. Accordingly, we focus on a dataset containing ground-truth VHT values alongside corresponding VHT measurements derived from virtual loop detectors for all $n=1570$ road segments, computed at $L=39$ time steps (corresponding to a 3-minute aggregation over the selected 2-hour interval) for each of the 101 simulated time series.



\subsection{Real data overview}\label{sec:athens-data}

The second dataset is based on the real-world pNEUMA experiment~(\cite{barmpounakis2020}), in which ten camera-equipped drones recorded traffic over approximately $1.3\:\text{km}^2$ of central Athens during morning peak periods in October and November 2018. Vehicle trajectories were reconstructed from the aerial footage across the monitored area. As in the simulation study, we assume that whenever a road segment is within a drone's field of view, the corresponding ground-truth traffic variable is available. This dataset therefore enables evaluation of the proposed framework under real-world urban traffic conditions.

\medskip Following the setup in Section~\ref{sec:method}, the monitored area is modeled as an undirected weighted graph using data collected over four recording days. On each day, the drones performed four consecutive half-hour flight sessions covering the morning peak from 08:30 to 10:30. Between consecutive recording sessions, the drones had to land, undergo battery replacement, and relaunch. As a result, only part of each half-hour period was effectively recorded, and the duration of the recorded portion varied from day to day. The footage of a given day covered between $393$ and $400$ road segments in total, but the set of observed segments also varied across sessions. We therefore preprocess the recordings to retain only the segments and time intervals that were observed consistently across all sessions and days. This yields a core of $n=327$ segments, forming a connected subgraph with $780$ edges, each linking two segments that meet at a common intersection. Traffic variables are aggregated over $90\,$s intervals. After intersecting the available intervals across the four days, each session contributes roughly the first $15$ minutes of its half-hour period, resulting in a total of $L=37$ time steps per day. The resulting common observation window spans from 08:30 to 10:15, with gaps between sessions corresponding to battery replacements. Although limited in size by machine-learning standards, this dataset provides a realistic test case for the proposed GP framework, which can operate effectively in data-scarce settings where data-intensive models such as deep neural networks may be difficult to train reliably.

\medskip In addition to the drone recordings, virtual loop detector measurements extracted from the captured trajectories are also available at the road segments. For every $90\,\mathrm{s}$ interval, each loop detector reports the traffic flow and the mean speed at its location, from which the corresponding segment-level VKT and VHT estimates are computed. As in Section~\ref{sec:data_sim}, we focus on VHT, as it provides a more direct indication of traffic congestion than VKT.

\subsection{Training procedure}

The available data in both case studies are divided into training and test sets, with training data originating from two complementary sensing sources. Historical loop-detector measurements provide spatially complete, although imperfect, VHT observations and are collected in the set $\mc{H}_{\text{LD}}$. Drone flights, on the other hand, provide more accurate VHT measurements collected in $\mc{H}_{\text{D}}$, but only over a subset of the spatio-temporal domain. These training data are used to construct the mean function of the Gaussian process in equation~\eqref{eq:trafficGP} and to learn the parameters of the kernel capturing the spatio-temporal correlations introduced in equations~\eqref{eq:sep_kernel}-\eqref{eq:temporal_kernel}. The resulting GP model is subsequently kept fixed and evaluated on previously unseen traffic realizations contained in the test set $\mc{H}_{\text{test}}$.

\medskip
Recall that the Gaussian process is defined over inputs of the form $\mathbf{x}=(v_i,t_j)$, where $v_i$ denotes a road segment and $t_j$ denotes a discrete time step. We assume that complete loop detector information is available for $N_{\text{LD}}$ days, represented by the $N_{\text{LD}}$ time series in $\mc{H}_{\text{LD}}$. In simulations, these time series come from repeated realizations of the traffic process, while in the real world they correspond to real-world loop detector measurements over multiple days. Furthermore, we assume that drone flights were conducted over $N_{\text{D}}$ days, resulting in the $N_{\text{D}}$ time series with observations for a subset of all possible spatio-temporal inputs. 

Since multiple measurements are available from loop detectors and  potentially drones for each road segment and each time step across various days, we define the prior mean function $m(\mathbf{x})$ in a nonparametric manner as an empirical average over the available observations at each spatio-temporal location. For the $\ell$-th realization in $\mc{H}_{\text{LD}}$, let $y_{\text{LD}}^{(\ell)}(\mathbf{x})$ denote the loop detector VHT measurement at time step and location $\mathbf{x}$. Similarly, let $y_{\text{D}}^{(\ell)}(\mathbf{x})$ denote the VHT value from drone observations at $\mathbf{x}$ in the $\ell$-th time series of drone flights $\mc{H}_{\text{D}}$.  As drone measurements are sparse, some inputs might appear once or multiple times in $\mc{H}_{\text{D}}$, while others might not appear at all. For simplicity, let $y_{\text{D}}^{(\ell)}(\mathbf{x})=0$ if an input $\mathbf{x}$ is not covered by drones in time series $\ell$. If $n(\mathbf{x})$ is the number of times $\mathbf{x}$ was covered by drones, the historical mean can be defined as
\begin{equation}
    m(\mathbf{x})=\frac{1}{N_{\text{LD}}+\beta n(\mathbf{x})}\left( \sum_{\ell\in\mc{H}_{\text{LD}}}y_{\text{LD}}^{(\ell)}(\mathbf{x}) + \beta \sum_{\ell\in\mc{H}_{\text{D}}}y_{\text{D}}^{(\ell)}(\mathbf{x})\right),
\end{equation}
where $\beta>0$ determines the relative weight assigned to the more accurate drone measurements. For $\beta=1$, all available measurements contribute equally, whereas $\beta>1$ assigns greater importance to drone observations. The function $m(\mathbf{x})$ is therefore interpreted as an empirical estimate of the recurrent traffic pattern at a given road segment and time of day. It is constructed exclusively from previously collected data and is kept fixed during evaluation.

Drone observations from $\mc{H}_{\text{D}}$ are also used to learn the parameters of the kernel $k(\mathbf{x},\mathbf{x'})$, which is intended to capture spatio-temporal dependencies that persist across different realizations of the traffic dynamics. When multiple measurements are available for the same input $\mathbf{x}$ across training days, we aggregate the repeated observations $y_D^{(\ell)}(\mathbf{x})$ according to
\begin{equation}\label{eq:avg_obs}
    \overline{y}(\mathbf{x})=\frac{1}{n(\mathbf{x})}\sum_{\ell\in\mc H_{\text{D}}}y_D^{(\ell)}(\mathbf{x}), \hspace{3mm} \text{for } n(\mathbf{x})\neq 0.
\end{equation}
This averaging step serves to suppress realization-specific fluctuations and measurement noise, thereby emphasizing traffic patterns that recur systematically across the training series. In particular, under the idealized assumption of independent realizations of a common process, averaging repeated observations reduces their variance while preserving the underlying spatial and temporal correlation structure. Consequently, the averaged observations in~\eqref{eq:avg_obs} form a smoothed training representation of the recurrent spatio-temporal dynamics and provide a more stable empirical baseline for identifying characteristic correlation scales.
Since estimating spatio-temporal correlations is computationally demanding and sensitive to noisy inputs, we rely solely on drone data for this purpose. The dataset $\mc{H}_{\text{D}}$ is inherently smaller than $\mc{H}_{\text{LD}}$, and can be subsampled if necessary to maintain computational efficiency. Furthermore, because drone coverage is sparse, the number of measurements $n(\mathbf{x})$ naturally varies across inputs, leading to non-uniform statistical precision in the averaged observations. For tractability, rather than modeling this heterogeneity explicitly, we represent the residual variability of the averaged training observations through a single effective Gaussian noise parameter $\sigma_n$. 

\medskip
The GP hyperparameters and the noise variance $\theta=\begin{bmatrix}
    \sigma_f, \nu_g, \kappa_g, \ell_{\mc T}, \sigma_n
\end{bmatrix}$ are estimated by maximizing the marginal log‑likelihood of the training data. Under the adopted working observation model $$\overline{y} \:\:|\:\: \mathbf{X},\theta\sim\mc N(\mathbf{m},\mathbf{K}_\theta),$$ where $\overline{\mathbf{y}}$ denotes the smoothed observations, $\mathbf{X}$ denotes the training inputs (i.e., road segments and time steps) corresponding to those observations, and $\mathbf{K}_\theta = \mathbf{K}_{\mathbf{XX}} + \sigma_n^2 \mathbf{I}_N$, the optimal parameters are obtained by maximizing  $L(\theta)=\log p(\mathbf{y} \mid \mathbf{X}, \theta)$, i.e.,
\begin{equation}
    \theta^*=\argmax_{\theta} \left(-\dfrac{1}{2}(\mathbf{y}-\mathbf{m})^T\mathbf{K}_\theta^{-1}(\mathbf{y}-\mathbf{m}) - \dfrac{1}{2}\log \det(\mathbf{K}_\theta)-\dfrac{N}{2}\log(2\pi)\right),
\end{equation}
where $N$ is the number of available observations. Training proceeds for a fixed number of gradient-based iterations, after which the optimized hyperparameters define the trained GP model used for posterior inference. The resulting hyperparameter values for both case studies are reported in Table~\ref{tab:appendix_gp} in Appendix~\ref{app:gp_hyperparameters}.

\medskip Note that this training procedure separates recurrent traffic structure from realization-specific information. Historical loop-detector measurements and drone observations are used to construct an informative prior mean, and repeated high-quality drone observations are averaged to expose persistent spatio-temporal dependencies and reduce the influence of day-specific stochastic fluctuations. At test time, no averaging across realizations is performed. Observations from the previously unseen traffic realization are incorporated directly through the GP posterior equations in Section 3, allowing the model to update the historical prior using the traffic conditions observed in real time.

\section{Results and discussion}\label{sec:results}

\begin{figure}[h!]
     \centering
     \includegraphics[width=0.538\textwidth]{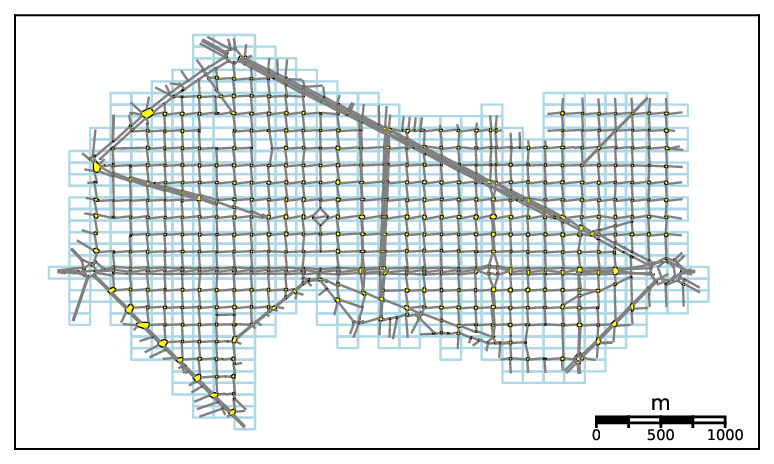}
     \includegraphics[width=0.452\textwidth]{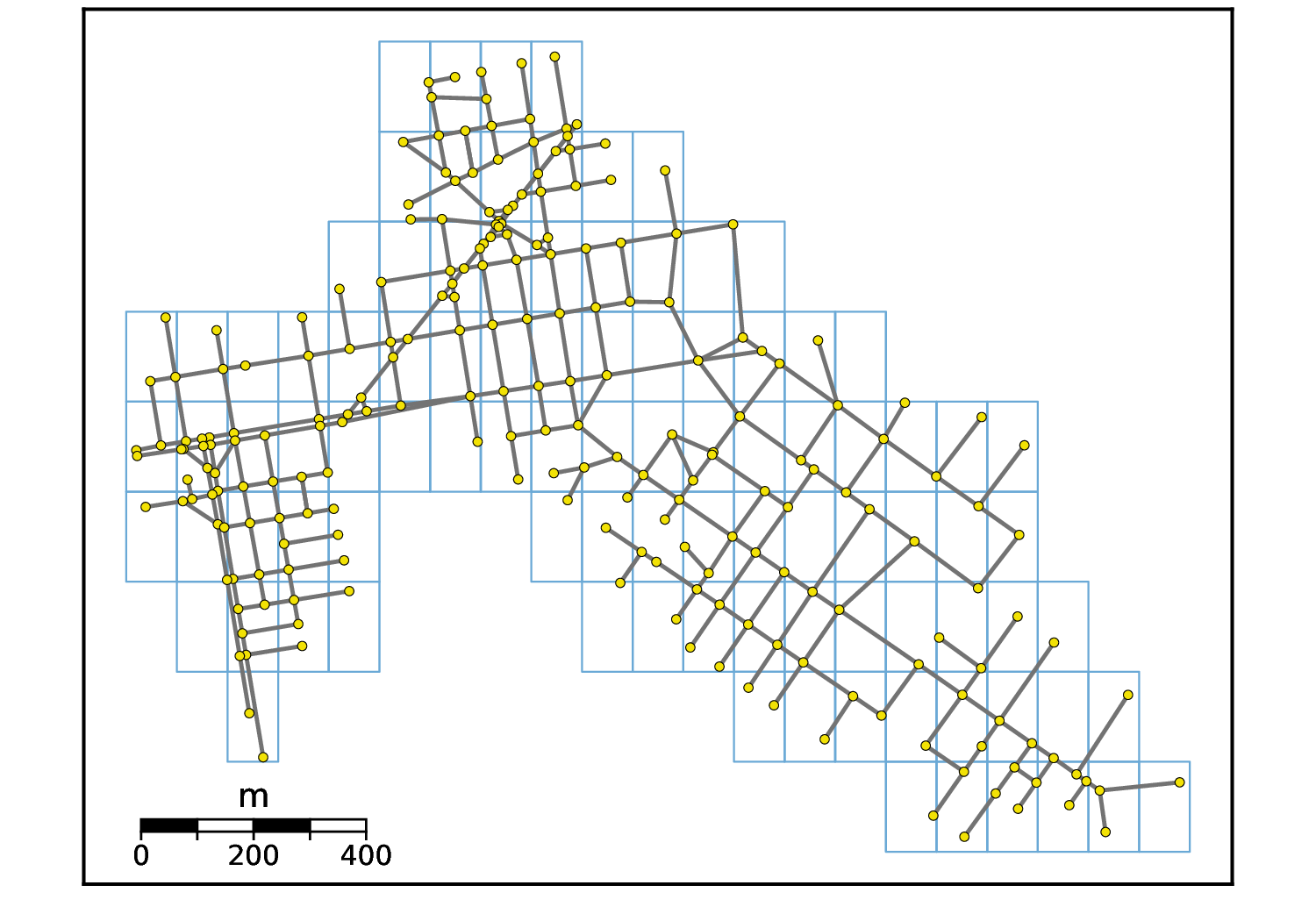}
     \caption{Traffic network in Barcelona (left) and Athens (right) divided into rectangular cells of $160\,\mathrm{m} \times 90\,\mathrm{m}$ representing the field of view of a single drone. The Athens network is shown rotated by $90^\circ$ for compactness.}
     \label{fig:network}
\end{figure}

Our approaches are tested using simulation data in the city of Barcelona and real-world data in the city of Athens. In both case studies, a fleet of $P$ drones is distributed in the urban area, and it is assumed that each drone can monitor an area of $160\,\mathrm{m} \times 90\,\mathrm{m}$ with its camera from an altitude of approximately $100\,\mathrm{m}$. Therefore, the urban areas are divided into rectangular subareas of the specified size, as shown in Figure \ref{fig:network}. At each time step, each of the $P$ drones is located in one of the cells of the grid, and a single cell is never monitored by two or more drones at the same time. We assume that a drone can observe all road segments that are (even partially) inside the cell under monitoring, so that it is capable of providing complete and accurate information about VHT in those road segments. We also assume that drones can change their positions by moving to one of the four adjacent cells in a way that they synchronously take the new positions at the beginning of the next time step. The interval between two consecutive time steps is long enough for a drone to take precise observations in the current cell and move to an adjacent cell.

\medskip
We investigate how the number of drones, which is directly related to the number of available observations, improves the estimates originating from loop detectors and influences the quality of the GP estimates. We conduct this analysis for two deployment scenarios: static drones (GP-S) and dynamic drones (GP-D). In the static case, drones are randomly distributed across the grid at the beginning of the experiment, with the probability of selecting each cell proportional to the number of road segments it covers. Their positions then remain fixed over time, so observations are always collected from the same subset of road segments. In the dynamic case, each drone can either move to one of the adjacent cells or remain at its current position at every time step. For a fair comparison, the drones end up in the same final positions as in the GP-S case, but during preceding time steps they can observe different parts of the network. Drone movements are selected randomly but coordinated centrally to avoid overlap, ensuring that no two drones monitor the same cell. Since some cells are more informative, the results depend on the stochastic distribution of the drones. Therefore, the outputs of each analysis are averaged over multiple runs.

\medskip
The Gaussian process parameters are first determined from the training data using the described procedure with $\beta=1$ and then fixed during the evaluation process. Each time series from the test set $\mc H_{\text{test}}$ is then used to evaluate the quality of the GP estimates on a day previously unseen. The results reported below are the average over all time series in $\mc H_{\text{test}}$. Using the GP parameters learned from loop detector data together with the observations collected by the drone fleet at the current time step $t$ and $k$ preceding time steps $t-1, t-2, \dots$, $t-k$, we can compute the posterior distributions of the traffic variables for all road segments. When the objective is to estimate the traffic variables at the current time step $t$, we refer to the task as \textit{traffic state estimation}. When the objective is instead to infer the traffic variables one step ahead, at time $t+1$, we refer to it as \textit{traffic state prediction}. 

\medskip
The performance of the GP is compared against a baseline relying solely on historical loop detector data from $\mc{H}_{\text{LD}}$, namely the error obtained when using the historical mean at the corresponding time step as the estimate for the current day (HM). In addition, we also consider a second benchmark that combines the historical mean with current observations (HMO): for all road segments covered by drones, the observed values are used directly, while for unobserved segments, the historical mean is used. 
The estimation error of each model for an arbitrary query point $\mathbf{x}_* = (v_*, t_*)$ is given by $(\hat{y}^{M}_*-y_*)^2$, where $y_*$ is the ground-truth value of the traffic parameter in question for a road segment $v_*$ at a time step $t_*$, and $\hat{y}^M_*$ is the estimate using the method $M\in\{\text{HM}, \text{HMO}, \text{GP-S}, \text{GP-D}\}$. To evaluate the estimation quality across the entire network of $n$ road segments at a fixed time step $t_*$ for a given method $M$, we use the relative root square error given by the formula
\begin{equation} \label{eq:errM}
    \operatorname{err}(M) = \sqrt{\dfrac{\sum\limits_{i=1}^n (\hat{y}^M_i-y_i)^2}{\sum\limits_{i=1}^n (\hat{y}^{\text{HM}}_i-y_i)^2}}
\end{equation}
Note that $\operatorname{err}(\text{HM})$ is always equal to 1. In other words, the metric defined in~\eqref{eq:errM} measures the relative improvement of each method compared to relying solely on the historical data provided by loop detectors.

\subsection{Case study: Simulation data}\label{sec:sim_data}

In the first case study, we use data generated using the Aimsun simulator described in Section~\ref{sec:data_sim}. Each time series covers a 2-hour interval, starting at the beginning of the demand peak, and is divided into 39 equidistant time steps, each separated by a 3-minute time interval. The training set consists of $N_{\text{LD}} = 30$ days of complete loop detector measurements in $\mc{H}_{\text{LD}}$, and $N_{\text{D}}=10$ days of accurate drone observations in $\mc{H}_{\text{D}}$ collected by a fleet of 100 drones moving randomly across the network. In total, 40 training time series are used, corresponding to demand scaling factors in $\{1, 1.20, 1.35, 1.50\}$. These scenarios remain relatively close to the historically calibrated demand, with an average demand scale of approximately $1.35$ across the training series. The remaining 18 time series featuring demands of $\{1.20, 1.35, 1.50\}$ are used for evaluation. Additionally, 43 time series with demand scales in $\{1.65, 1.80\}$ are included in the test set $\mc{H}_{\text{test}}$ to evaluate model performance under conditions of greater traffic congestion. All computations are implemented using the GPyTorch library (\cite{gardner2018gpytorch}). The GP model is trained for 500 iterations, requiring approximately 80~s on an NVIDIA RTX 4090 GPU. Inference of the GP model takes less than $10\,\mathrm{ms}$, which enables real-time execution.

\medskip
To provide a clear view of absolute traffic conditions across time, Figure~\ref{fig:barcelona_timeseries} shows the VHT for a representative test sequence from $\mathcal{H}_{\text{test}}$ under a demand scale of $1.20$, considering a fixed fleet of 40 static drones. We compare the ground-truth (GT) VHT against raw loop detector (LD) measurements, the historical mean (HM) derived from $\mathcal{H}_{\text{LD}}$, and the static Gaussian Process (GP-S) estimates across two representative road segments: a highly congested link (Link 588) and a less congested link (Link 1009). As shown, raw loop detector measurements consistently underestimate ground-truth VHT, which leads to biased historical baseline estimates. Incorporating drone observations within the GP framework substantially mitigates this offset, resulting in a closer tracking of ground-truth fluctuations. While Figure~\ref{fig:barcelona_timeseries} highlights GP performance in absolute units for a single time series, the remainder of this paper evaluates performance improvements across multiple sequences in $\mathcal{H}_{\text{test}}$, reporting estimation errors relative to the historical baseline.

\begin{figure}[tbh]
    \centering
    \begin{subcaptionblock}[t]
    {0.47\textwidth}
        \centering
        \resizebox{0.98\linewidth}{!}{
\begin{tikzpicture}

\definecolor{lightgray}{RGB}{211,211,211}
\definecolor{lightgray204}{RGB}{204,204,204}
\definecolor{purple}{RGB}{128,0,128}

\begin{axis}[
legend cell align={left},
legend style={
  fill opacity=0.8,
  draw opacity=1,
  text opacity=1,
  at={(0.03,0.03)},
  anchor=south west,
  font=\footnotesize,
  legend columns=2,
  draw=lightgray204
},
tick align=outside,
tick pos=left,
title={Congested link 588},
x grid style={lightgray},
xlabel={Time step},
xmajorgrids,
xmin=0, xmax=39.3,
xtick style={color=black},
y grid style={lightgray},
ylabel={VHT [veh\(\displaystyle \cdot\)s]},
ylabel style={font=\footnotesize},
ymajorgrids,
ymin=-0.0476614586074126, ymax=1.00089063075567,
ytick style={color=black}
]
\addplot [semithick, black, mark=*,  mark size=2, mark options={solid}]
table {%
2 0.410300303574912
3 0.618866076316804
4 0.578963682001897
5 0.81506859033257
6 0.659237840825072
7 0.587859936163989
8 0.687367696661709
9 0.688361797322147
10 0.582809894349369
11 0.891286113205414
12 0.666191516520938
13 0.386742771207193
14 0.616233021757783
15 0.731237760887552
16 0.47569252388782
17 0.546959674519783
18 0.953229172148252
19 0.69998142796837
20 0.654111102547201
21 0.457852305975687
22 0.706284302733864
23 0.660031005611951
24 0.397173116114254
25 0.662176337016818
26 0.797579658872273
27 0.621702522408405
28 0.60005606239541
29 0.873543776590197
30 0.88966268673526
31 0.531109731952748
32 0.776581433592319
33 0.750201368093613
34 0.607590972215788
35 0.551779770136038
36 0.292534190687969
37 0.00851782421019379
38 0
};
\addlegendentry{GT}
\addplot [semithick, purple, dashed, mark=x, mark size=2.2, mark options={solid}]
table {%
2 0.372873276339006
3 0.430402815414639
4 0.451762001258552
5 0.505941105609646
6 0.461802552813871
7 0.469220124697205
8 0.524122765596355
9 0.471926114895177
10 0.468525279233344
11 0.556969640828812
12 0.476691125207843
13 0.359628388113168
14 0.458937219759737
15 0.494648613775868
16 0.420037671458238
17 0.451465943405669
18 0.573833578990475
19 0.491848834062223
20 0.487662210222419
21 0.372976384193573
22 0.515438232129155
23 0.460367794013338
24 0.355626085774692
25 0.474659262475868
26 0.515800933951047
27 0.440270176742789
28 0.434043089151336
29 0.599834187663915
30 0.49855030606173
31 0.370846787571357
32 0.515817653610242
33 0.503325078357482
34 0.465486139159335
35 0.426545765440679
36 0.201519747665359
37 0.0082053960198327
38 0
};
\addlegendentry{LD}
\addplot [semithick, red, dashed, mark=triangle*, mark size=2, mark options={solid}]
table {%
2 0.48844139277935
3 0.493391141295433
4 0.490240484476089
5 0.516347274184227
6 0.501268431544304
7 0.506747826933861
8 0.506990656256676
9 0.508145168423653
10 0.53599601984024
11 0.530275940895081
12 0.52145254611969
13 0.530301332473755
14 0.520456910133362
15 0.538140133023262
16 0.55849626660347
17 0.535318359732628
18 0.555562153458595
19 0.539770692586899
20 0.541553735733032
21 0.533969208598137
22 0.560957193374634
23 0.55853059887886
24 0.549766078591347
25 0.527171641588211
26 0.529538422822952
27 0.565220788121223
28 0.551617875695229
29 0.550984025001526
30 0.504991874098778
31 0.513419061899185
32 0.566607311367989
33 0.550444364547729
34 0.545861005783081
35 0.425265625119209
36 0.371310532093048
37 0.199001312255859
38 0.158393941819668
};
\addlegendentry{HM}
\addplot [semithick, blue, mark=square*, mark size=1.65, mark options={solid}]
table {%
2 0.533182114362717
3 0.551502004265785
4 0.561093360185623
5 0.61678509414196
6 0.586479395627975
7 0.57010643184185
8 0.589728310704231
9 0.586270406842232
10 0.607639342546463
11 0.637867033481598
12 0.602160662412643
13 0.576788708567619
14 0.597801804542542
15 0.63067638874054
16 0.607901930809021
17 0.580508351325989
18 0.729308277368546
19 0.60857418179512
20 0.603676065802574
21 0.575798615813255
22 0.639454811811447
23 0.618310108780861
24 0.659501776099205
25 0.600404530763626
26 0.621771544218063
27 0.622851267457008
28 0.612205311655998
29 0.653524294495583
30 0.618090078234673
31 0.549753248691559
32 0.637058392167091
33 0.635943308472633
34 0.599283188581467
35 0.479233771562576
36 0.396214842796326
37 0.197899729013443
38 0.165690712630749
};
\addlegendentry{GP-S}
\end{axis}

\end{tikzpicture}}
        \caption{}
        \label{fig:barcelona_timeseries_high}
    \end{subcaptionblock}
    \begin{subcaptionblock}[t]{0.47\textwidth}
        \centering
        \resizebox{0.98\linewidth}{!}{
\begin{tikzpicture}

\definecolor{lightgray}{RGB}{211,211,211}
\definecolor{lightgray204}{RGB}{204,204,204}
\definecolor{purple}{RGB}{128,0,128}

\begin{axis}[
legend cell align={left},
legend style={
  fill opacity=0.8,
  draw opacity=1,
  text opacity=1,
  at={(0.03,0.03)},
  anchor=south west,
  font=\footnotesize,
  legend columns=2,
  draw=lightgray204
},
tick align=outside,
tick pos=left,
title={Uncongested link 1009},
x grid style={lightgray},
xlabel={Time step},
xmajorgrids,
xmin=0, xmax=39.3,
xtick style={color=black},
scaled y ticks=false,
ytick={0, 0.05, 0.1, 0.15},
yticklabel style={/pgf/number format/fixed},
y grid style={lightgray},
ylabel={VHT [veh\(\displaystyle \cdot\)s]},
ylabel style={font=\footnotesize},
ymajorgrids,
ymin=-0.002, ymax=0.16,
ytick style={color=black}
]
\addplot [semithick, black, mark=*, mark size=2, mark options={solid}]
table {%
2 0.0933797951870946
3 0.0816537346106733
4 0.0846047299896202
5 0.103174103122849
6 0.121248033558794
7 0.105328584607171
8 0.101612243042727
9 0.0923976047166684
10 0.106465989527796
11 0.0966719468027044
12 0.0826977812214982
13 0.066496598628195
14 0.135653337237257
15 0.0715663713482867
16 0.0590751828174814
17 0.13156614986789
18 0.0741429959470727
19 0.103792812988677
20 0.155333340841529
21 0.0995112277050655
22 0.0858827469656717
23 0.150975819907528
24 0.148813881538877
25 0.0530128168916758
26 0.0656226027605478
27 0.138628852204822
28 0.136001811568953
29 0.139017505299866
30 0.0648780238675373
31 0.110236573004283
32 0.0993674918063142
33 0.120808370989444
34 0.102139859265293
35 0.11982334805597
36 0.0585265514595795
37 0.0120606687087328
38 0.0031512945672446
};
\addlegendentry{GT}
\addplot [semithick, purple, dashed, mark=x, mark size=2.2, mark options={solid}]
table {%
2 0.0751603866751561
3 0.0361271404147277
4 0.0715601541407339
5 0.0611115593286743
6 0.0917507721319697
7 0.0659972308366456
8 0.0781462716792554
9 0.0824998752054253
10 0.0685703118698299
11 0.0573402747222029
12 0.070095334018491
13 0.0640344973156307
14 0.0752739727555579
15 0.0565661784010789
16 0.0505639753549416
17 0.0809977538604736
18 0.0582041678593508
19 0.0847855168794988
20 0.0795762264959603
21 0.0758777415461777
22 0.0568798017730959
23 0.0867796923086398
24 0.102316483419755
25 0.0470021676178687
26 0.0455222431609972
27 0.0858532509830804
28 0.0706679529883169
29 0.0921197145120507
30 0.0595109252945284
31 0.0665073607452396
32 0.0647384197665065
33 0.0744013847823956
34 0.064721996578047
35 0.0603620891249782
36 0.0295928876689156
37 0.0115173125688272
38 0.00307296617181762
};
\addlegendentry{LD}
\addplot [semithick, red, dashed, mark=triangle*, mark size=2, mark options={solid}]
table {%
2 0.0640526786446571
3 0.0713607296347618
4 0.0697277188301086
5 0.0712519362568855
6 0.0724096670746803
7 0.0783620476722717
8 0.071384459733963
9 0.0700314044952393
10 0.0653800889849663
11 0.0701120942831039
12 0.0769657269120216
13 0.074150912463665
14 0.0669263601303101
15 0.07459606975317
16 0.0852267891168594
17 0.0680861175060272
18 0.0726239755749702
19 0.0758342519402504
20 0.0772752016782761
21 0.0764288008213043
22 0.077419213950634
23 0.0643607750535011
24 0.0727367699146271
25 0.0726859644055367
26 0.0666741579771042
27 0.0764445140957832
28 0.0707580596208572
29 0.069103054702282
30 0.0732032805681229
31 0.0741494446992874
32 0.0801959335803986
33 0.077911265194416
34 0.0713245868682861
35 0.0556857101619244
36 0.0177288912236691
37 0.0169700514525175
38 0.00752756465226412
};
\addlegendentry{HM}
\addplot [semithick, blue, mark=square*, mark size=1.65, mark options={solid}]
table {%
2 0.0895483791828156
3 0.0871274173259735
4 0.0906998664140701
5 0.100189320743084
6 0.113085098564625
7 0.104064024984837
8 0.0997208207845688
9 0.0949123650789261
10 0.100313529372215
11 0.09166319668293
12 0.0964580029249191
13 0.104679748415947
14 0.113328516483307
15 0.0954142212867737
16 0.0950999334454536
17 0.105295799672604
18 0.10283187776804
19 0.113386727869511
20 0.125311493873596
21 0.108112849295139
22 0.0945806577801704
23 0.115356355905533
24 0.125898838043213
25 0.0985523536801338
26 0.0938219577074051
27 0.128491967916489
28 0.118493445217609
29 0.113060139119625
30 0.0890887752175331
31 0.10975157469511
32 0.111030146479607
33 0.113506689667702
34 0.106579892337322
35 0.0925061777234077
36 0.0378092974424362
37 0.0259698089212179
38 0.00838884990662336
};
\addlegendentry{GP-S}
\end{axis}

\end{tikzpicture}}
        \caption{}
        \label{fig:barcelona_timeseries_low}
    \end{subcaptionblock}
    \caption{
    Time series of VHT from time step 2 to 38 for a representative evaluation sequence with demand scale $1.20$. The figure compares ground-truth (GT) VHT against raw loop detector (LD) measurements, historical mean (HM) baseline values, and static Gaussian Process (GP-S) estimates for (a) a highly congested road segment (Link 588) and (b) an uncongested road segment (Link 1009). The GP-S method uses observations from the current and two previous time steps collected by 40 static drones to compute VHT estimates. The results are averaged over 200 simulation runs with different drone placements.}
    \label{fig:barcelona_timeseries}
\end{figure}

\medskip
We deploy a fleet of $P\in\{10, 20, 40, 60, 80, 100, 150, 200\}$ randomly distributed drones over the Barcelona grid, and report VHT estimation results averaged over 100 runs. For the traffic state estimation task, VHT values are estimated for each road segment at the current time step $t$ using partial observations from the current and two previous time steps. We first evaluate our methods on the 18 lower-demand time series in $\mc{H}_{\text{test}}$ and show the distribution of the relative estimation error with respect to the number of active drones in Figure~\ref{fig:mean_curve_estimation_low}. The procedure is then repeated for the remaining 43 time series with largest demand in $\mc{H}_{\text{test}}$, with the corresponding results shown in Figure~\ref{fig:mean_curve_estimation_high}. We compare four different methods: historical mean of loop detector data (HM), historical mean with observations (HMO), Gaussian process with fixed drones (static case, GP-S), and Gaussian process with moving drones (dynamic case, GP-D), and summarize the numerical values in Table~\ref{tab:results_barcelona}.


\begin{figure}[t]
    \centering
    \begin{subcaptionblock}[t]
    {0.47\textwidth}
        \centering
        \resizebox{0.98\linewidth}{!}{
\begin{tikzpicture}

\definecolor{green}{RGB}{0,128,0}
\definecolor{lightgray}{RGB}{211,211,211}
\definecolor{lightgray204}{RGB}{204,204,204}
\definecolor{orange}{RGB}{255,165,0}

\begin{axis}[
title={Estimation under lower demand},
legend cell align={left},
legend style={
  fill opacity=0.8,
  draw opacity=1,
  text opacity=1,
  at={(0.03,0.03)},
  anchor=south west,
  draw=lightgray204
},
tick align=outside,
tick pos=left,
x grid style={lightgray},
xlabel={Active drones},
xmajorgrids,
xmin=0, xmax=210,
xtick style={color=black},
y grid style={lightgray},
ylabel={Estimation error},
ymajorgrids,
ymin=0.285700615413194, ymax=1.05,
ytick style={color=black}
]
\addplot [semithick, red, mark=*, mark size=3, mark options={solid}]
table {%
10 1.00016
20 1.00016
40 1.00016
60 1.00016
80 1.00016
100 1.00016
150 1.00016
200 1.00016
};
\addlegendentry{HM}
\addplot [semithick, orange, mark=*, mark size=3, mark options={solid}]
table {%
10 0.971504795279807
20 0.944696046437996
40 0.89093343035691
60 0.840517441890187
80 0.790461991278778
100 0.743798807749973
150 0.631975925344197
200 0.537898546278811
};
\addlegendentry{HMO}
\addplot [semithick, blue, mark=*, mark size=3, mark options={solid}]
table {%
10 0.952305826043413
20 0.921452017146788
40 0.859951382356222
60 0.804557060187817
80 0.751529746882048
100 0.702785230691889
150 0.589732806687374
200 0.497834615810831
};
\addlegendentry{GP-S}
\addplot [semithick, green, mark=*, mark size=3, mark options={solid}]
table {%
10 0.937259963971451
20 0.893025895967756
40 0.810038740767636
60 0.7394616153072
80 0.673188060094067
100 0.616261856496498
150 0.492329947139106
200 0.397294990232646
};
\addlegendentry{GP-D}
\end{axis}

\end{tikzpicture}}
        \caption{}
        \label{fig:mean_curve_estimation_low}
    \end{subcaptionblock}
    \begin{subcaptionblock}[t]{0.47\textwidth}
        \centering
        \resizebox{0.98\linewidth}{!}{
\begin{tikzpicture}

\definecolor{green}{RGB}{0,128,0}
\definecolor{lightgray}{RGB}{211,211,211}
\definecolor{lightgray204}{RGB}{204,204,204}
\definecolor{orange}{RGB}{255,165,0}

\begin{axis}[
title={Estimation under higher demand},
legend cell align={left},
legend style={
  fill opacity=0.8,
  draw opacity=1,
  text opacity=1,
  at={(0.03,0.03)},
  anchor=south west,
  draw=lightgray204
},
tick align=outside,
tick pos=left,
x grid style={lightgray},
xlabel={Active drones},
xmajorgrids,
xmin=0, xmax=210,
xtick style={color=black},
y grid style={lightgray},
ylabel={Estimation error},
ymajorgrids,
ymin=0.273787330516253, ymax=1.05,
ytick style={color=black}
]
\addplot [semithick, red, mark=*, mark size=3, mark options={solid}]
table {%
10 1.00016
20 1.00016
40 1.00016
60 1.00016
80 1.00016
100 1.00016
150 1.00016
200 1.00016
};
\addlegendentry{HM}
\addplot [semithick, orange, mark=*, mark size=3, mark options={solid}]
table {%
10 0.974063200388644
20 0.948307691150129
40 0.898992697902932
60 0.851812924532705
80 0.806237261388723
100 0.762014224312246
150 0.659061795826118
200 0.562945659819589
};
\addlegendentry{HMO}
\addplot [semithick, blue, mark=*, mark size=3, mark options={solid}]
table {%
10 0.959033836261703
20 0.924263209198042
40 0.859657750977934
60 0.80038920389552
80 0.745553255097662
100 0.694448550157799
150 0.580436675727216
200 0.480745967804017
};
\addlegendentry{GP-S}
\addplot [semithick, green, mark=*, mark size=3, mark options={solid}]
table {%
10 0.941908601030934
20 0.892006065459314
40 0.802451738144726
60 0.723140322850513
80 0.653148384997989
100 0.591378344198928
150 0.461793945249028
200 0.359232236993394
};
\addlegendentry{GP-D}
\end{axis}

\end{tikzpicture}}
        \caption{}
        \label{fig:mean_curve_estimation_high}
    \end{subcaptionblock}
    \caption{
    Mean estimation error versus numbers of active drones in case of lower (a) and greater congestion (b). Partial observations from $t=24,25,26$ are used to estimate the full-graph values at $t=26$. Each curve represents the average performance over the test time sequences, where for each sequence the results are first averaged over 100 simulation runs with different initial drone placements, and then averaged across all available sequences.}
    \label{fig:mean_curve_estimation}
\end{figure}

\medskip
 By using current drone observations in combination with the historical mean, the estimation error can be substantially reduced. When the GP framework is applied, the error is reduced further, achieving noticeably better performance in both the static and dynamic cases. In the dynamic scenario, where drones monitor connected cells in the neighborhood during consecutive time steps by making random movements to adjacent cells, the framework achieves the lowest relative estimation error of approximately 0.40 for a fleet of 200 drones. The improvements offered by the GPs are particularly noticeable when traffic demand scales are significantly larger than the historically calibrated baseline, becoming also more pronounced relative to the HMO baseline. Specifically, under high demand, the relative estimation error of around 0.36 for 200 drones in the GP-D case represents a 36\% reduction compared to the HMO curve, outperforming the 26\% improvement observed for lower demands. This shows that GPs are better suited to adapt to drastic variations in traffic conditions. Finally, it can be seen that comparable performance is achieved by using 200 drones without the GP framework (HMO curve), 150 static drones with the GP, or just 100 dynamic drones with the GP. Hence, the integration of GPs can noticeably reduce the fleet size required to achieve a certain estimation precision.

 \begin{table}[t]
\centering
\footnotesize
\setlength{\tabcolsep}{5pt}
\caption{Estimation and prediction error
on the Barcelona simulation network for different numbers of active drones, reported relative to the HM baseline (which is equal to one in every cell by construction). HM is therefore omitted from the table, along with HMO in the prediction task, as they coincide in this case. Here, $t\in[\cdot]$ is the observation window and $t_*$ is the target time step at which the variable is inferred.}
\label{tab:results_barcelona}
\vspace{2mm}
\begin{tabular}{c | l l *{8}{S[table-format=1.3]}}
\toprule
\multicolumn{1}{c}{} & & & \multicolumn{8}{c}{\textbf{Active drones}} \\
\cmidrule(lr){4-11}
\multicolumn{1}{c}{} & & \textbf{Method} & {10} & {20} & {40} & {60} & {80} & {100} & {150} & {200} \\
\midrule
\multirow{6}{*}{\rotatebox[origin=c]{90}{\textbf{Estimate}}}
 & \multirow{3}{*}{\shortstack[l]{$t\in[24,26]$\\$t_*=26$\\low dem.}}
   & HMO & 0.972 & 0.945 & 0.891 & 0.841 & 0.790 & 0.744 & 0.632 & 0.538 \\
 & & GP-S & 0.952 & 0.921 & 0.860 & 0.805 & 0.752 & 0.703 & 0.590 & 0.498 \\
 & & GP-D & 0.937 & 0.893 & 0.810 & 0.739 & 0.673 & 0.616 & 0.492 & 0.397 \\[4pt]
 & \multirow{3}{*}{\shortstack[l]{$t\in[24,26]$\\$t_*=26$\\high dem.}}
   & HMO & 0.974 & 0.948 & 0.899 & 0.852 & 0.806 & 0.762 & 0.659 & 0.563 \\
 & & GP-S & 0.959 & 0.924 & 0.860 & 0.800 & 0.746 & 0.694 & 0.580 & 0.481 \\
 & & GP-D & 0.942 & 0.892 & 0.802 & 0.723 & 0.653 & 0.591 & 0.462 & 0.359 \\
\midrule
\multirow{4}{*}{\rotatebox[origin=c]{90}{\textbf{Predict}}}
 & \multirow{2}{*}{\shortstack[l]{$t\in[4,6]$\\$t_*=7$}}
   & GP-S & 0.966 & 0.943 & 0.902 & 0.864 & 0.830 & 0.798 & 0.732 & 0.681 \\
 & & GP-D & 0.956 & 0.924 & 0.868 & 0.819 & 0.777 & 0.741 & 0.672 & 0.624 \\[4pt]
 & \multirow{2}{*}{\shortstack[l]{$t\in[29,31]$\\$t_*=32$}}
   & GP-S & 0.966 & 0.941 & 0.893 & 0.850 & 0.812 & 0.776 & 0.702 & 0.643 \\
 & & GP-D & 0.954 & 0.918 & 0.853 & 0.799 & 0.751 & 0.711 & 0.634 & 0.582 \\
\bottomrule
\end{tabular}
\end{table}

\medskip
In case of the traffic state prediction task, we evaluate the ability of the previously introduced methods to forecast VHT values one time step ahead of the latest available observations. As in the estimation task, a fleet of $P\in\{10, 20, 40, 60, 80, 100, 150, 200\}$ randomly distributed drones is deployed on the grid, and partial observations collected at the three preceding time steps are used to predict VHT values for all road segments at the next time step. We report results for two different time windows within the simulation horizon, as shown in Figure~\ref{fig:mean_curve_pred}. Each curve represents the average error over 100 simulation runs with different initial drone placements and across all 61 sequences in $\mc{H}_{\text{test}}$.

\begin{figure}[t]
    \centering
    \begin{subcaptionblock}[t]
    {0.47\textwidth}
        \centering
        \resizebox{0.98\linewidth}{!}{
\begin{tikzpicture}

\definecolor{green}{RGB}{0,128,0}
\definecolor{lightgray}{RGB}{211,211,211}
\definecolor{lightgray204}{RGB}{204,204,204}
\definecolor{orange}{RGB}{255,165,0}

\begin{axis}[
title={Prediction for the early time window},
legend cell align={left},
legend style={
  fill opacity=0.8,
  draw opacity=1,
  text opacity=1,
  at={(0.03,0.03)},
  anchor=south west,
  draw=lightgray204
},
tick align=outside,
tick pos=left,
x grid style={lightgray},
xlabel={Active drones},
xmajorgrids,
xmin=0, xmax=210,
xtick style={color=black},
y grid style={lightgray},
ylabel={Prediction error},
ymajorgrids,
ymin=0.508702648567432, ymax=1.05,
ytick style={color=black}
]
\addplot [semithick, orange, mark=*, mark size=3, mark options={solid}]
table {%
10 0.99984
20 0.99984
40 0.99984
60 0.99984
80 0.99984
100 0.99984
150 0.99984
200 0.99984
};
\addlegendentry{HMO=HM}
\addplot [semithick, blue, mark=*, mark size=3, mark options={solid}]
table {%
10 0.966112474002636
20 0.943175217511033
40 0.902079272396329
60 0.863852748341927
80 0.829521570289617
100 0.797985481116723
150 0.732229725777366
200 0.680746433250547
};
\addlegendentry{GP-S}
\addplot [semithick, green, mark=*, mark size=3, mark options={solid}]
table {%
10 0.955539210151743
20 0.923612497392146
40 0.868123491262323
60 0.819017868074942
80 0.77734247372027
100 0.741357915502568
150 0.671700294512442
200 0.624366625969644
};
\addlegendentry{GP-D}
\end{axis}

\end{tikzpicture}}
        \caption{}
        \label{fig:mean_curve_pred_a}
    \end{subcaptionblock}
    \begin{subcaptionblock}[t]{0.47\textwidth}
        \centering
        \resizebox{0.98\linewidth}{!}{
\begin{tikzpicture}

\definecolor{green}{RGB}{0,128,0}
\definecolor{lightgray}{RGB}{211,211,211}
\definecolor{lightgray204}{RGB}{204,204,204}
\definecolor{orange}{RGB}{255,165,0}

\begin{axis}[
title={Prediction for the later time window},
legend cell align={left},
legend style={
  fill opacity=0.8,
  draw opacity=1,
  text opacity=1,
  at={(0.03,0.03)},
  anchor=south west,
  draw=lightgray204
},
tick align=outside,
tick pos=left,
x grid style={lightgray},
xlabel={Active drones},
xmajorgrids,
xmin=0, xmax=210,
xtick style={color=black},
y grid style={lightgray},
ylabel={Prediction error},
ymajorgrids,
ymin=0.459072507390712, ymax=1.05,
ytick style={color=black}
]
\addplot [semithick, orange, mark=*, mark size=3, mark options={solid}]
table {%
10 0.99984
20 0.99984
40 0.99984
60 0.99984
80 0.99984
100 0.99984
150 0.99984
200 0.99984
};
\addlegendentry{HMO=HM}
\addplot [semithick, blue, mark=*, mark size=3, mark options={solid}]
table {%
10 0.965761114643296
20 0.940656111749925
40 0.892543447579895
60 0.850326167892445
80 0.811843957730224
100 0.775513738052175
150 0.701721548374497
200 0.643336162471939
};
\addlegendentry{GP-S}
\addplot [semithick, green, mark=*, mark size=3, mark options={solid}]
table {%
10 0.953544042564623
20 0.917970766661136
40 0.853132546576861
60 0.798637968353942
80 0.751382527307034
100 0.710742190982522
150 0.634101765064639
200 0.581949889869399
};
\addlegendentry{GP-D}
\end{axis}

\end{tikzpicture}}
        \caption{}
        \label{fig:mean_curve_pred_b}
    \end{subcaptionblock}
    \caption{
    Mean prediction error versus numbers of active drones in the Barcelona network. In (a), partial observations from $t=4,5,6$ are used to predict the full-graph values at $t=7$, while in (b), observations from $t=29,30,31$ are used to predict the values at $t=32$.}
    \label{fig:mean_curve_pred}
\end{figure}

\medskip
Since the prediction target lies in the future with respect to the latest measurements, no drone observations are available at the target time step itself. Thus, the historical mean with observations (HMO) method reduces to the plain historical mean (HM) method, resulting in the two curves overlapping in both plots. While the GP-based methods outperform the historical mean methods, the magnitude of the error depends on the specific forecasting time window. In Figure \ref{fig:mean_curve_pred_a}, all methods exhibit a slightly larger relative error compared to Figure \ref{fig:mean_curve_pred_b}. Nonetheless, in both experiments, the relative ordering of the methods is preserved, and the GP provides a moderate improvement even with a small number of drones. As the number of available drones increases, both the static and dynamic GP variants achieve solid performance, although the absolute error reduction is smaller compared to the estimation experiment. Overall, these results indicate that the proposed spatio-temporal GP can be used for short-horizon forecasting of network-wide traffic parameters from sparse drone observations, and that its advantage over the historical mean is preserved across different phases of the daily demand profile. Finally, Table~\ref{tab:results_barcelona} also summarizes the exact numerical values of the relative errors for the prediction task.

\subsection{Case study: Real data}\label{sec:real_data}

In the second case study, we evaluate the framework on the real data collected in the city of Athens, as described in Section~\ref{sec:athens-data}. The data cover the morning peak interval starting at 08:30 and include $n=327$ road segments over $L=37$ equidistant time steps, with 90-second interval between two consecutive time steps. The dataset comprises four recording days, three of which form the training set while the remaining day is held out as the test set $\mc{H}_{\text{test}}$. From two of the three training days, we use both complete loop detector measurements and partial observations collected by a fleet of 20 randomly moving drones to construct the historical mean. The drone observations from the remaining training day are then used only to learn the kernel parameters capturing the spatio-temporal correlations. This separation avoids fitting the kernel parameters to the data already used for the mean in the case when the training dataset is highly limited. The GP model is trained for 1000 iterations, taking less than 20~s on an NVIDIA RTX 4090 GPU. Its inference time is below $10\,\mathrm{ms}$.

We deploy a fleet of $P\in\{2, 5, 10, 15, 20, 25, 30, 35, 40\}$ randomly distributed drones on the grid in Athens, and report the results of the VHT estimates averaged over 1000 runs. For the traffic state estimation task, VHT values are estimated for every road segment at the target time step using partial drone observations collected at the current and three preceding time steps. As in the Barcelona study, the error of each method is reported relative to the baseline given by the historical mean of loop detector data (HM), which uses the historical mean computed from $\mc{H}_{\text{LD}}$ as the prediction for the test day. Figure~\ref{fig:athens_estimation} reports the relative estimation error against the number of active drones for an early window in (a) and a later window in (b), further supported by exact numerical values summarized in Table~\ref{tab:results_athens}. The results closely mirror those obtained in the Barcelona simulation study, with the four methods preserving the same relative ordering across all fleet sizes. The historical mean with observations (HMO) already lowers the error well below the HM baseline, and the Gaussian process reduces it further in both the static case (GP-S) and the dynamic case (GP-D). GP-D, in which drones monitor connected cells over consecutive time steps, attains the lowest error across the entire range of fleet sizes. For the early window, a fleet of 40 drones brings the GP-D error down to roughly one third of the HM baseline, and the later window exhibits the same qualitative trend.

\begin{figure}[t]
    \centering
    \begin{subcaptionblock}[t]{0.47\textwidth}
        \centering
        \resizebox{0.98\linewidth}{!}{
\begin{tikzpicture}

\definecolor{green}{RGB}{0,128,0}
\definecolor{lightgray}{RGB}{211,211,211}
\definecolor{lightgray204}{RGB}{204,204,204}
\definecolor{orange}{RGB}{255,165,0}

\begin{axis}[
title={Estimation for the early time window},
legend cell align={left},
legend style={
  fill opacity=0.8,
  draw opacity=1,
  text opacity=1,
  at={(0.03,0.03)},
  anchor=south west,
  draw=lightgray204
},
tick align=outside,
tick pos=left,
x grid style={lightgray},
xlabel={Active drones},
xmajorgrids,
xmin=0, xmax=42.5,
xtick style={color=black},
y grid style={lightgray},
ylabel={Estimation error},
ymajorgrids,
ymin=0.25, ymax=1.04,
ytick style={color=black}
]
\addplot [semithick, red, mark=*, mark size=3, mark options={solid}]
table {%
2 1
5 1
10 1
15 1
20 1
25 1
30 1
35 1
40 1
};
\addlegendentry{HM}
\addplot [semithick, orange, mark=*, mark size=3, mark options={solid}]
table {%
2 0.972110180426546
5 0.932206391366767
10 0.861958598934203
15 0.798618950800284
20 0.731498679037459
25 0.663365823479498
30 0.603587016959265
35 0.540179119309331
40 0.481213779458893
};
\addlegendentry{HMO}
\addplot [semithick, blue, mark=*, mark size=3, mark options={solid}]
table {%
2 0.920702154640768
5 0.879219022163152
10 0.806874245007812
15 0.742845131326723
20 0.67556679501075
25 0.608655646080056
30 0.551137539573277
35 0.490513917023268
40 0.435014578554987
};
\addlegendentry{GP-S}
\addplot [semithick, green, mark=*, mark size=3, mark options={solid}]
table {%
2 0.906617581255112
5 0.846801957944644
10 0.750423265887564
15 0.666519876855818
20 0.586036810000174
25 0.516740703221948
30 0.456373968004854
35 0.398374304648676
40 0.344556233214818
};
\addlegendentry{GP-D}
\end{axis}

\end{tikzpicture}}
        \caption{}
        \label{fig:athens_estimation_a}
    \end{subcaptionblock}
    \begin{subcaptionblock}[t]{0.47\textwidth}
        \centering
        \resizebox{0.98\linewidth}{!}{
\begin{tikzpicture}

\definecolor{green}{RGB}{0,128,0}
\definecolor{lightgray}{RGB}{211,211,211}
\definecolor{lightgray204}{RGB}{204,204,204}
\definecolor{orange}{RGB}{255,165,0}

\begin{axis}[
title={Estimation for the later time window},
legend cell align={left},
legend style={
  fill opacity=0.8,
  draw opacity=1,
  text opacity=1,
  at={(0.03,0.03)},
  anchor=south west,
  draw=lightgray204
},
tick align=outside,
tick pos=left,
x grid style={lightgray},
xlabel={Active drones},
xmajorgrids,
xmin=0, xmax=42.5,
xtick style={color=black},
y grid style={lightgray},
ylabel={Estimation error},
ymajorgrids,
ymin=0.25, ymax=1.04,
ytick style={color=black}
]
\addplot [semithick, red, mark=*, mark size=3, mark options={solid}]
table {%
2 1
5 1
10 1
15 1
20 1
25 1
30 1
35 1
40 1
};
\addlegendentry{HM}
\addplot [semithick, orange, mark=*, mark size=3, mark options={solid}]
table {%
2 0.971855127181048
5 0.929593038542706
10 0.860777811180799
15 0.800555099194478
20 0.729328566976051
25 0.669018144984411
30 0.599716931241202
35 0.5400862247228
40 0.479551566322472
};
\addlegendentry{HMO}
\addplot [semithick, blue, mark=*, mark size=3, mark options={solid}]
table {%
2 0.885213487538453
5 0.845870966147012
10 0.781813761263904
15 0.724857470381653
20 0.658214689971797
25 0.602288218873335
30 0.538578225597742
35 0.481678998373318
40 0.427203122198186
};
\addlegendentry{GP-S}
\addplot [semithick, green, mark=*, mark size=3, mark options={solid}]
table {%
2 0.874755231783788
5 0.820428635320804
10 0.739316176449043
15 0.671454966735606
20 0.596097271938839
25 0.534319452800233
30 0.474833206867741
35 0.419308446276583
40 0.366375117192302
};
\addlegendentry{GP-D}
\end{axis}

\end{tikzpicture}}
        \caption{}
        \label{fig:athens_estimation_b}
    \end{subcaptionblock}
    \caption{
    Mean estimation error versus number of active drones in the Athens network. Partial observations from time steps (a) $t=4,5,6,7$ and (b) $t=22,23,24,25$ are used to estimate the full-graph VHT at $t=7$ and $t=25$, respectively. Each curve is averaged over 1000 runs with different initial drone placements on the single test day.}
    \label{fig:athens_estimation}
\end{figure}

\medskip The same figure can also be interpreted at a fixed error level rather than at a fixed fleet size, comparing the number of active drones each method needs to reach a given improvement. For the early window in Figure~\ref{fig:athens_estimation_a}, a relative error of $0.6$ is attained by GP-D with about 20 drones, by GP-S with about 25 to 26 drones, and by HMO only with about 30 drones. The later window in Figure~\ref{fig:athens_estimation_b} shows the same behavior, with GP-D consistently reaching any target error using fewer drones than GP-S and HMO. GP-D therefore reaches the same accuracy with the smallest fleet across both windows, which is consistent with its lower error at every fleet size and indicates a more efficient use of the available observations.

\medskip
For the traffic state prediction task, the fleet sizes used in the estimation task are retained. The objective is now to forecast network-wide VHT one time step ahead of the latest available measurements, using observations collected over the three preceding time steps. Figure~\ref{fig:athens_prediction} shows the results for an early window in (a) and a later window in (b). Analogously to the Barcelona study in Section~\ref{sec:sim_data}, since no observations are available at the future target step, HMO coincides with HM and remains at unit relative error. The Gaussian process methods outperform HM across all fleet sizes, although the improvement is more moderate than in the estimation task. GP-S and GP-D follow each other closely, with GP-D keeping a small advantage. The relative improvement grows with the size of the drone fleet, and for the largest fleet, the early window reaches a lower error than the later one. The prediction rows of Table~\ref{tab:results_athens} report the exact corresponding errors for both time windows.
\begin{table}[t]
\centering
\footnotesize
\setlength{\tabcolsep}{5pt}
\caption{Estimation and prediction error in the Athens network for different numbers of active drones, reported relative to the HM baseline (which is equal to one in every cell by construction). HM is therefore omitted from the table, along with HMO in the prediction task, as they coincide in this case. Here, $t\in[\cdot]$ is the observation window and $t_*$ is the target time step at which the variable is inferred.}
\label{tab:results_athens}
\vspace{2mm}
\begin{tabular}{c | l l *{9}{S[table-format=1.3]}}
\toprule
\multicolumn{1}{c}{} & & & \multicolumn{9}{c}{\textbf{Active drones}} \\
\cmidrule(lr){4-12}
\multicolumn{1}{c}{} & & \textbf{Method} & {2} & {5} & {10} & {15} & {20} & {25} & {30} & {35} & {40} \\
\midrule
\multirow{6}{*}{\rotatebox[origin=c]{90}{\textbf{Estimate}}}
 & \multirow{3}{*}{\shortstack[l]{$t\in[4,7]$\\$t_*=7$}}
   & HMO & 0.972 & 0.932 & 0.862 & 0.799 & 0.731 & 0.663 & 0.604 & 0.540 & 0.481 \\
 & & GP-S & 0.921 & 0.879 & 0.807 & 0.743 & 0.676 & 0.609 & 0.551 & 0.491 & 0.435 \\
 & & GP-D & 0.907 & 0.847 & 0.750 & 0.667 & 0.586 & 0.517 & 0.456 & 0.398 & 0.345 \\[4pt]
 & \multirow{3}{*}{\shortstack[l]{$t\in[22,25]$\\$t_*=25$}}
   & HMO & 0.972 & 0.930 & 0.861 & 0.801 & 0.729 & 0.669 & 0.600 & 0.540 & 0.480 \\
 & & GP-S & 0.885 & 0.846 & 0.782 & 0.725 & 0.658 & 0.602 & 0.539 & 0.482 & 0.427 \\
 & & GP-D & 0.875 & 0.820 & 0.739 & 0.671 & 0.596 & 0.534 & 0.475 & 0.419 & 0.366 \\
\midrule
\multirow{4}{*}{\rotatebox[origin=c]{90}{\textbf{Predict}}}
 & \multirow{2}{*}{\shortstack[l]{$t\in[12,14]$\\$t_*=15$}}
   & GP-S & 0.934 & 0.918 & 0.892 & 0.863 & 0.840 & 0.814 & 0.798 & 0.779 & 0.763 \\
 & & GP-D & 0.926 & 0.901 & 0.862 & 0.822 & 0.795 & 0.768 & 0.751 & 0.739 & 0.724 \\[4pt]
 & \multirow{2}{*}{\shortstack[l]{$t\in[22,24]$\\$t_*=25$}}
   & GP-S & 0.899 & 0.882 & 0.852 & 0.824 & 0.804 & 0.780 & 0.760 & 0.743 & 0.729 \\
 & & GP-D & 0.894 & 0.872 & 0.836 & 0.806 & 0.782 & 0.761 & 0.742 & 0.729 & 0.716 \\
\bottomrule
\end{tabular}
\end{table}
\begin{figure}[t]
    \centering
    \begin{subcaptionblock}[t]{0.47\textwidth}
        \centering
        \resizebox{0.98\linewidth}{!}{
\begin{tikzpicture}

\definecolor{green}{RGB}{0,128,0}
\definecolor{lightgray}{RGB}{211,211,211}
\definecolor{lightgray204}{RGB}{204,204,204}
\definecolor{orange}{RGB}{255,165,0}

\begin{axis}[
title={Prediction for the early time window},
legend cell align={left},
legend style={
  fill opacity=0.8,
  draw opacity=1,
  text opacity=1,
  at={(0.03,0.03)},
  anchor=south west,
  draw=lightgray204
},
tick align=outside,
tick pos=left,
x grid style={lightgray},
xlabel={Active drones},
xmajorgrids,
xmin=0, xmax=45,
xtick style={color=black},
y grid style={lightgray},
ylabel={Prediction error},
ymajorgrids,
ymin=0.65, ymax=1.02,
ytick style={color=black}
]
\addplot [semithick, orange, mark=*, mark size=3, mark options={solid}]
table {%
2 1
5 1
10 1
15 1
20 1
25 1
30 1
35 1
40 1
};
\addlegendentry{HMO=HM}
\addplot [semithick, blue, mark=*, mark size=3, mark options={solid}]
table {%
2 0.93437766858779
5 0.918072122054866
10 0.892407343472251
15 0.862772683907491
20 0.8399232723155
25 0.813762008612224
30 0.798196448057548
35 0.779276612166635
40 0.7634110919521
};
\addlegendentry{GP-S}
\addplot [semithick, green, mark=*, mark size=3, mark options={solid}]
table {%
2 0.925821647748848
5 0.900759410478649
10 0.861926762453935
15 0.822037365673829
20 0.795060269669217
25 0.767818978848868
30 0.751272365754363
35 0.738577662581614
40 0.723575607088117
};
\addlegendentry{GP-D}
\end{axis}

\end{tikzpicture}}
        \caption{}
        \label{fig:athens_prediction_a}
    \end{subcaptionblock}
    \begin{subcaptionblock}[t]{0.47\textwidth}
        \centering
        \resizebox{0.98\linewidth}{!}{
\begin{tikzpicture}

\definecolor{green}{RGB}{0,128,0}
\definecolor{lightgray}{RGB}{211,211,211}
\definecolor{lightgray204}{RGB}{204,204,204}
\definecolor{orange}{RGB}{255,165,0}

\begin{axis}[
title={Prediction for the later time window},
legend cell align={left},
legend style={
  fill opacity=0.8,
  draw opacity=1,
  text opacity=1,
  at={(0.03,0.03)},
  anchor=south west,
  draw=lightgray204
},
tick align=outside,
tick pos=left,
x grid style={lightgray},
xlabel={Active drones},
xmajorgrids,
xmin=0, xmax=42.5,
xtick style={color=black},
y grid style={lightgray},
ylabel={Prediction error},
ymajorgrids,
ymin=0.65, ymax=1.02,
ytick style={color=black}
]
\addplot [semithick, orange, mark=*, mark size=3, mark options={solid}]
table {%
2 1
5 1
10 1
15 1
20 1
25 1
30 1
35 1
40 1
};
\addlegendentry{HMO=HM}
\addplot [semithick, blue, mark=*, mark size=3, mark options={solid}]
table {%
2 0.898750097690181
5 0.881840608383824
10 0.852359412445909
15 0.82421548706759
20 0.803852898874336
25 0.780314198656389
30 0.759598633572841
35 0.742512962459452
40 0.728806221271858
};
\addlegendentry{GP-S}
\addplot [semithick, green, mark=*, mark size=3, mark options={solid}]
table {%
2 0.893848139534951
5 0.871569925729535
10 0.836108862581397
15 0.805559312523772
20 0.782404423459455
25 0.760880890811043
30 0.742121938914738
35 0.728743837817887
40 0.715850031088108
};
\addlegendentry{GP-D}
\end{axis}

\end{tikzpicture}}
        \caption{}
        \label{fig:athens_prediction_b}
    \end{subcaptionblock}
    \caption{
    Mean prediction error versus number of active drones in the Athens network. Partial observations from (a) $t=12,13,14$ and (b) $t=22,23,24$ are used to predict the full-graph VHT one step ahead at $t=15$ and $t=25$, respectively. Each curve is averaged over 1000 runs with different initial drone placements on the single test day.}
    \label{fig:athens_prediction}
\end{figure}

\medskip
Overall, the experiments on the Athens network confirm the behavior observed throughout this study. Both GP-S and GP-D lower the relative error below the HM baseline across all fleet sizes, in the estimation as well as in the one-step-ahead prediction setting. In the estimation task, GP-D attains the lowest error, as the mobility of the fleet provides observations over connected cells across consecutive time steps. In the prediction task, this advantage largely vanishes and GP-S and GP-D perform comparably. In all configurations, the relative error decreases monotonically with the number of deployed drones. The performance improvement is more modest than in the controlled simulation setting, which is attributable to the limited number of recording days available for training and the higher measurement noise inherent to real data. 

\subsection{Uncertainty estimation}\label{sec:uncertaity}

A central advantage of the Gaussian process framework is its ability to provide a principled representation of the predictive uncertainty in addition to point estimates. In our setting, this means that the model returns not only an estimate of the traffic variable at each road segment, but also an associated measure of confidence in that estimate. This uncertainty is inherently observation-dependent and evolves over time, reflecting both the decreasing relevance of older measurements and the changing spatial coverage induced by the movement of the drone fleet.

\begin{figure}[t]
    \centering
    \begin{subcaptionblock}[t]{0.48\textwidth}
        \centering
        \includegraphics[width=\textwidth]{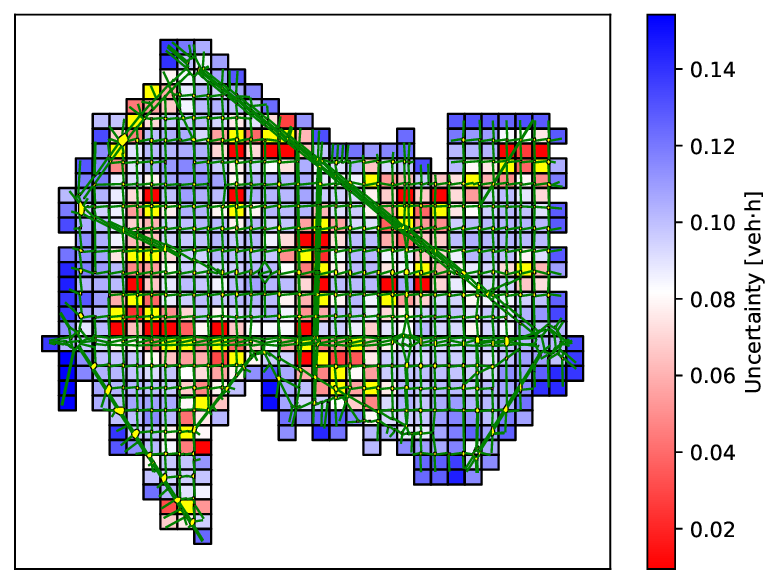}
        \caption{Barcelona with 40 drones at time step $t=26$}
        \label{fig:uncertainty_map_barcelona}
    \end{subcaptionblock}
    \begin{subcaptionblock}[t]{0.48\textwidth}
        \centering
        \includegraphics[width=\textwidth]{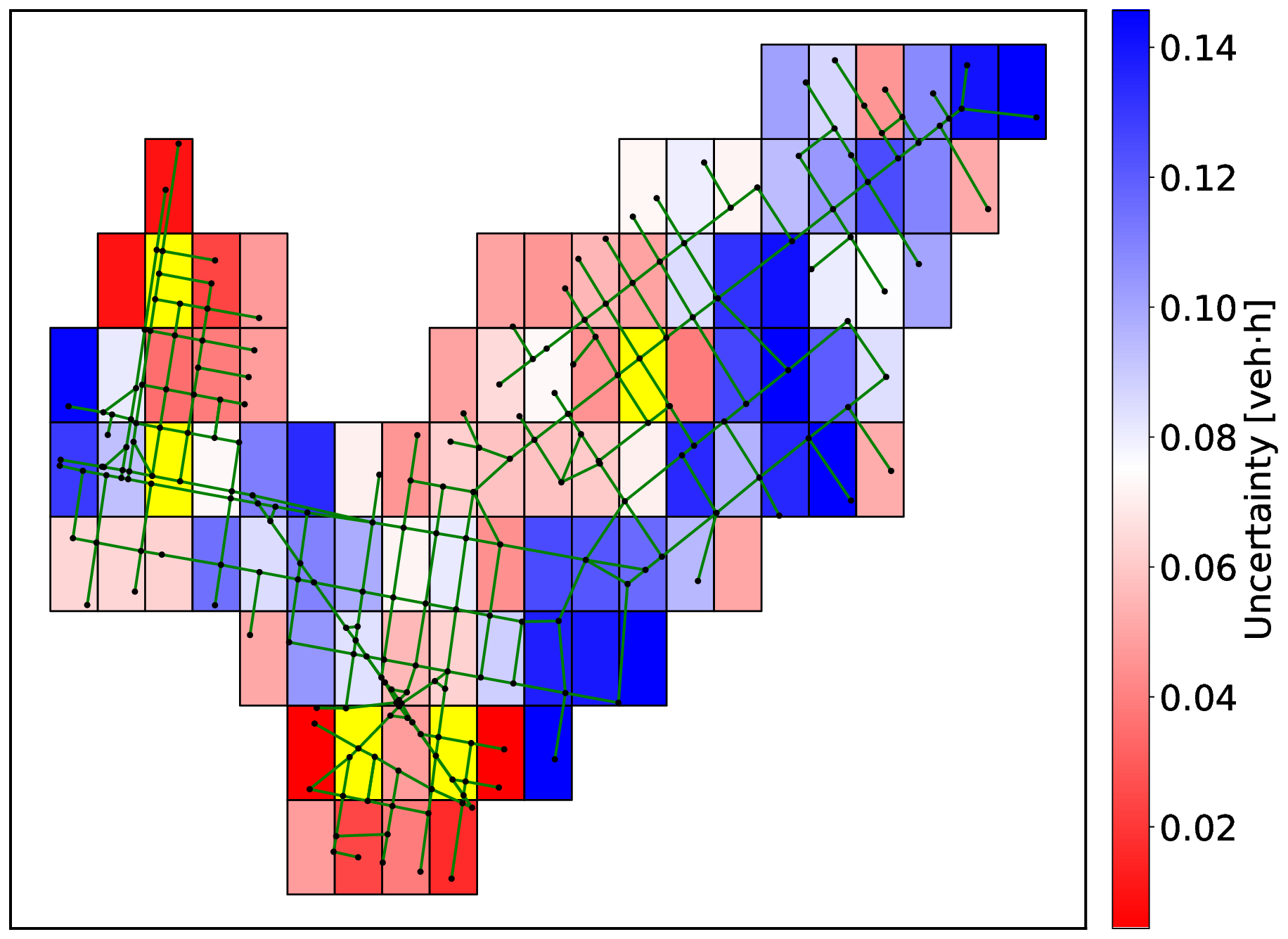}
        \caption{Athens with 5 drones at time step $t=7$}
        \label{fig:uncertainty_map_athens}
    \end{subcaptionblock}
    \caption{
    Uncertainty maps for a fixed time step and a single simulation run in the moving drones case (GP-D). Yellow cells correspond to the current drone positions. Red cells have smaller uncertainty, while blue ones have large uncertainty.}
    \label{fig:uncertainty_maps}
\end{figure}

Given that drone observations provide precise estimates of the current VHT values for the covered road segments, the uncertainties of all observed segments are very low. Beyond these directly observed locations, the uncertainty exhibits a clear spatio-temporal character. Spatially, segments in cells neighboring an observation have lower uncertainty than those located further away. Temporally, in the dynamic monitoring case, recently visited cells exhibit lower uncertainty than the rest of the network, where no recent measurements are available. This behavior is illustrated in Figure \ref{fig:uncertainty_maps} for the dynamic drones case, which shows the uncertainty maps for a fixed time step and a single simulation run of a randomly selected sequence from $\mc H_{\text{test}}$. Specifically, Figure \ref{fig:uncertainty_map_barcelona} corresponds to a fleet of 40 drones in Barcelona, while Figure \ref{fig:uncertainty_map_athens} corresponds to 5 drones in the Athens case study.

To evaluate the quality of the uncertainty estimates, we report two standard prediction interval metrics: the Prediction Interval Coverage Probability (PICP) and the Mean Prediction Interval Width (MPIW)~(\cite{khosravi2011review}). For a nominal confidence level $1-\alpha$, the GP posterior at each road segment $v_i$ induces the prediction interval $\left[\hat{y}_i - z_{1-\alpha/2}\sigma_i,\; \hat{y}_i + z_{1-\alpha/2}\sigma_i\right]$, where $\hat{y}_i$ and $\sigma_i$ denote the corresponding posterior mean and standard deviation given by \eqref{eq:post_mean} and \eqref{eq:post_cov}, and $z_{1-\alpha/2} = \Phi^{-1}(1-\alpha/2)$ is the corresponding quantile of the standard normal distribution, where $\Phi$ denotes its cumulative distribution function. The PICP is the fraction of ground-truth values contained in this interval,
\begin{equation}
    \mathrm{PICP} = \frac{1}{n}\sum_{i=1}^{n} \mathbf{1}\!\left\{\hat{y}_i - z_{1-\alpha/2}\sigma_i \le y_i \le \hat{y}_i + z_{1-\alpha/2}\sigma_i\right\},
    \label{eq:picp}
\end{equation}
where $\mathbf{1}\{\cdot\}$ denotes the indicator function and $n$ is the number of road segments. The MPIW measures the average width of these intervals,
\begin{equation}
    \mathrm{MPIW} = \frac{1}{n}\sum_{i=1}^{n} 2 z_{1-\alpha/2}\sigma_i.
    \label{eq:mpiw}
\end{equation}
Thus, PICP evaluates the empirical coverage of the uncertainty intervals, while MPIW characterizes their sharpness. Ideally, PICP should be close to the nominal confidence level $1-\alpha$, with narrower intervals indicating more informative uncertainty estimates.

\medskip In all experiments, we use a $95\%$ confidence level, i.e., $\alpha = 0.05$ and $z_{1-\alpha/2} = z_{0.975} \approx 1.96$. Table~\ref{tab:uncertainty} reports the PICP and MPIW for both case studies, evaluated over all road segments and over the unobserved segments alone. This separation matters because observed segments carry very low predictive variance, so their intervals are narrow and contain the ground truth. Their inclusion raises the PICP and lowers the MPIW, which makes the all-segment values an optimistic summary of the uncertainty quality. The unobserved segments give a stricter measure, since these are the locations the model must infer. In Barcelona, the PICP is around $97.5\%$ over both all and unobserved segments, with an MPIW of $0.31\,\mathrm{veh\cdot h}$ over all segments and $0.36\,\mathrm{veh\cdot h}$ over unobserved ones. In Athens, the PICP is $94.70\%$ over all segments and $93.23\%$ over the unobserved ones, both close to the nominal $95\%$ level, while the MPIW widens from $0.27$ to $0.34\,\mathrm{veh\cdot h}$ once the tightly constrained observed segments are excluded. The empirical coverage stays close to the nominal $95\%$ level, indicating that the posterior variances provide a reliable quantification of the estimation uncertainty.

\begin{table}[h]
  \centering
  \caption{Uncertainty quality at the \qty{95}{\percent} nominal confidence level, reported over all road segments and over the unobserved segments only. PICP near \qty{95}{\percent} indicates reliable coverage, and smaller MPIW indicates sharper intervals.}
  \label{tab:uncertainty}
  \sisetup{table-format=2.2}
  \begin{tabular}{l S S S S}
    \toprule
    \multirow{2}{*}{Metric} & \multicolumn{2}{c}{Barcelona} & \multicolumn{2}{c}{Athens} \\
    \cmidrule(lr){2-3}\cmidrule(lr){4-5}
     & {All} & {Unobserved} & {All} & {Unobserved} \\
    \midrule
    PICP [\unit{\percent}]  & 97.52 & 97.48 & 94.70 & 93.23 \\
    MPIW [\unit{\veh\hour}] & 0.31  & 0.36  & 0.27  & 0.34  \\
    \bottomrule
  \end{tabular}
\end{table}

\medskip These uncertainty estimates can also be exploited to improve trajectory planning and decision-making. A straightforward strategy is to guide the drones at each time step toward the cells with the highest total uncertainty, defined as the sum of predictive uncertainties over all road segments within a cell, rather than allowing them to move at random. The resulting Gaussian process with uncertainty-aware movements (GP-U) approach is compared with the GP-S and GP-D approaches in Figure \ref{fig:gpu} for the Barcelona and Athens case studies. As in the previous analysis, drones in the GP-S and GP-D cases are deployed to the same cells in the final time step, randomly selected with probabilities proportional to the number of road segments in each cell. In contrast, in the GP-U approach, after initialization, the drones follow uncertainty-guided trajectories and are not constrained to reach predetermined final positions. The results are averaged over all available time sequences in $\mc H_{\text{test}}$, using 100 runs for Barcelona and 1000 runs for Athens, with varied initial drone placements.

\begin{figure}[t]
    \centering
    \begin{subcaptionblock}[t]
    {0.475\textwidth}
        \centering
        \resizebox{0.98\linewidth}{!}{
\begin{tikzpicture}

\definecolor{green}{RGB}{0,128,0}
\definecolor{lightgray}{RGB}{211,211,211}
\definecolor{lightgray204}{RGB}{204,204,204}
\definecolor{purple}{RGB}{128,0,128}

\begin{axis}[
title={Comparison of GP methods in Barcelona},
title style={font=\small},
legend cell align={left},
legend style={fill opacity=0.8, draw opacity=1, text opacity=1, draw=lightgray204},
tick align=outside,
tick pos=left,
x grid style={lightgray},
xlabel={Active drones},
xmajorgrids,
xmin=0, xmax=209.5,
xtick style={color=black},
y grid style={lightgray},
ylabel={Estimation error},
ymajorgrids,
ymin=-0.01, ymax=1.04,
ytick style={color=black}
]
\addplot [semithick, blue, mark=*, mark size=3, mark options={solid}]
table {%
10 0.956359407906927
20 0.922705914433878
40 0.860256756858693
60 0.801156050579956
80 0.746780114969155
100 0.696935546644069
150 0.583148200779775
200 0.485995868949777
};
\addlegendentry{GP-S}
\addplot [semithick, green, mark=*, mark size=3, mark options={solid}]
table {%
10 0.939783070118249
20 0.891617048696717
40 0.805349343522229
60 0.727336332467039
80 0.658461005244942
100 0.598212601436471
150 0.470769994405722
200 0.370702873060982
};
\addlegendentry{GP-D}
\addplot [semithick, purple, mark=*, mark size=3, mark options={solid}]
table {%
10 0.923377009838992
20 0.861264920815115
40 0.756720882942781
60 0.667408692273057
80 0.595375977711867
100 0.533376386642098
150 0.412586933420108
200 0.325522613438322
};
\addlegendentry{GP-U}
\end{axis}

\end{tikzpicture}}
        \caption{}
        \label{fig:gpu_barcelona}
    \end{subcaptionblock}
    \begin{subcaptionblock}[t]{0.47\textwidth}
        \centering
        \resizebox{0.98\linewidth}{!}{
\begin{tikzpicture}

\definecolor{green}{RGB}{0,128,0}
\definecolor{lightgray}{RGB}{211,211,211}
\definecolor{lightgray204}{RGB}{204,204,204}
\definecolor{purple}{RGB}{128,0,128}

\begin{axis}[
title={Comparison of GP methods in Athens},
title style={font=\small},
legend cell align={left},
legend style={fill opacity=0.8, draw opacity=1, text opacity=1, draw=lightgray204},
tick align=outside,
tick pos=left,
x grid style={lightgray},
xlabel={Active drones},
xmajorgrids,
xmin=0, xmax=52.5,
xtick style={color=black},
y grid style={lightgray},
ylabel={Estimation error},
ymajorgrids,
ymin=-0.01, ymax=1.04,
ytick style={color=black}
]
\addplot [semithick, blue, mark=*, mark size=3, mark options={solid}]
table {%
2 0.921536676851049
5 0.877598883946391
10 0.810023758750623
15 0.739318239804846
20 0.677991049401546
25 0.61242216103508
30 0.5505937824328
35 0.48841485257283
40 0.432149831977266
45 0.375506671004309
50 0.315905739018924
};
\addlegendentry{GP-S}
\addplot [semithick, green, mark=*, mark size=3, mark options={solid}]
table {%
2 0.907185096279189
5 0.845774755653434
10 0.750699924221442
15 0.664420670151643
20 0.591471988729287
25 0.519300590122267
30 0.453761079140878
35 0.396124097145418
40 0.343476601998985
45 0.294047369434328
50 0.246969297616772
};
\addlegendentry{GP-D}
\addplot [semithick, purple, mark=*, mark size=3, mark options={solid}]
table {%
2 0.897052064757324
5 0.830419497349682
10 0.728974016023458
15 0.636923541899185
20 0.558326750264607
25 0.4850745803235
30 0.421191325634479
35 0.358698812636001
40 0.301631103276153
45 0.250049347473461
50 0.206053336123048
};
\addlegendentry{GP-U}
\end{axis}

\end{tikzpicture}}
        \caption{}
        \label{fig:gpu_athens}
    \end{subcaptionblock}
    \caption{
    Mean estimation error versus numbers of active drones for different GP approaches. (a) Barcelona case study, where partial observations from time steps $t=24,25,26$ are used to estimate the full-graph values at $t=26$. (b) Athens case study, where partial observations from time steps $t=4,5,6,7$ are used to estimate the full-graph values at $t=7$.}
    \label{fig:gpu}
\end{figure}

The experiment in Figure \ref{fig:gpu} provides a proof-of-concept for using GP posterior uncertainty as a planning signal. Even with this simple greedy rule, which looks only one step ahead, the uncertainty-guided movements yield a clear improvement over the random GP-S and GP-D strategies across the range of fleet sizes considered. A potential limitation of such a myopic heuristic is that its advantage may diminish as the number of drones approaches the total number of cells and the network becomes almost fully observed, since little uncertainty then remains to be exploited. These results nonetheless highlight the value of uncertainty-aware planning and motivate future work on more elaborate control algorithms that look beyond a single step ahead.

\section{Conclusions}\label{sec:con}

In this paper, we demonstrated the potential of drones as mobile sensors for urban traffic monitoring. Drone observations complement fixed loop detector data and, due to their higher precision, can directly improve traffic state estimation. However, as traffic variables across the urban network are not independent random variables, but rather exhibit underlying spatio-temporal correlations, we introduced a spatio-temporal GP framework that combines partial drone observations with loop detector measurements to estimate current traffic states and provide short-term predictions.

\medskip The proposed model integrates a graph-based spatial kernel that accounts for the structure of the road network with a temporal kernel that captures how traffic evolves over time. We evaluated the framework on both simulated and real-world data, using Barcelona's road network under different demand levels and real traffic measurements from Athens. The results show that our method outperforms historical mean approaches, even against the optimistic baseline in which loop detectors are available in all road segments. For a given drone fleet size, the Gaussian process framework yields more accurate traffic state estimates than the baseline methods. Alternatively, for a target level of estimation accuracy, the proposed approach can achieve comparable performance with a smaller number of drones, indicating a more efficient use of the available mobile sensing resources. These improvements are particularly noticeable when drone mobility is exploited, rather than treating them as static sensors. Furthermore, it was shown that Gaussian processes can be used to improve short-term predictions of traffic variables starting from available partial observations.

\medskip Beyond point estimates, the Gaussian process model provides predictive uncertainty, enabling a probabilistic characterization of the estimated traffic state. We showed that uncertainty is lower near monitored cells and higher in less observed parts of the network. Moreover, even a simple greedy uncertainty-aware movement strategy further improves estimation performance. This highlights the value of mobile sensing and uncertainty-aware planning, and motivates future work on multi-step control algorithms that optimize drone trajectories by explicitly reducing predictive uncertainty over time.

\section*{Acknowledgements}
This work was supported as a part of NCCR Automation, a National Centre of Competence in Research, funded by the Swiss National Science Foundation (grant number 51NF40\_225155).

\bibliography{references.bib}
\bibliographystyle{apalike} 

\clearpage

\appendix
\section{Appendix: Gaussian Process Hyperparameters}\label{app:gp_hyperparameters}

Table~\ref{tab:appendix_gp} summarizes the optimized hyperparameters of the Gaussian process models obtained via marginal log-likelihood maximization for the Barcelona and Athens case studies.

\begin{table}[h]
  \centering
  \caption{Optimized GP kernel and noise hyperparameters across both case studies.}
  \label{tab:appendix_gp}
  \begin{tabular}{l c c c}
    \toprule
    \textbf{Description} & \textbf{Parameter} & \textbf{Barcelona} & \textbf{Athens} \\
    \midrule
    Output scale & $\sigma_f^2$ & 0.5146 & 0.7467 \\
    Graph Matérn smoothness & $\nu_g$ & 4.0689 & 0.6921 \\
    Graph Matérn scale & $\kappa_g$ & 3.3969 & 2.1924 \\
    Temporal lengthscale & $\ell_{\mc T}$ & 0.0793 & 0.0584 \\
    Measurement noise & $\sigma_n^2$ & 0.0040 & 0.0018 \\
    \bottomrule
  \end{tabular}
\end{table}

\end{document}